\documentclass[11pt,a4paper]{article}
\usepackage{jheppub,amsmath,  amssymb,slashed,url,bm,textgreek,upgreek,footmisc}
\usepackage{graphicx}
\usepackage{epstopdf}

\usepackage{tikz}
\usetikzlibrary{decorations.pathreplacing, decorations.markings,calc,shapes.misc,decorations.pathmorphing,patterns.meta, math}

\newcommand\U{\text{U}}

\newcommand\SU{\text{SU}}

\newcommand{\beq}{\begin{equation}}
\newcommand{\eeq}{\end{equation}}
 
\newcommand{\bea}{\begin{eqnarray}}
\newcommand{\ea}{\end{eqnarray}}

\def\spin{{\mathrm{spin}}}

\def\tt{{\mathfrak t}}

\def\be{\begin{equation}}
\def\ee{\end{equation}}

\def\tilde{\widetilde}

\def\d{{\mathrm d}}

\def\[{\bigl [}

\def\]{\bigr ]}

\def\H{{\mathcal H}}

\def\tilde{\widetilde}

\def\bar{\overline}

\font\teneurm=eurm10 \font\seveneurm=eurm7  \font\fiveeurm=eurm5
\newfam\eurmfam
\textfont\eurmfam=\teneurm \scriptfont\eurmfam=\seveneurm
\scriptscriptfont\eurmfam=\fiveeurm

\font\teneusm=eusm10 \font\seveneusm=eusm7 \font\fiveeusm=eusm5
\newfam\eusmfam
\textfont\eusmfam=\teneusm \scriptfont\eusmfam=\seveneusm
\scriptscriptfont\eusmfam=\fiveeusm

\font\tencmmib=cmmib10 \skewchar\tencmmib='177
\font\sevencmmib=cmmib7 \skewchar\sevencmmib='177
\font\fivecmmib=cmmib5 \skewchar\fivecmmib='177
\newfam\cmmibfam
\textfont\cmmibfam=\tencmmib \scriptfont\cmmibfam=\sevencmmib
\scriptscriptfont\cmmibfam=\fivecmmib

\def\i{{\mathrm i}}

\def\SU{{\mathrm{SU}}}

\def\SU{{\mathrm{SU}}}

\title{Spin-Statistics for Black Hole Microstates\vspace{-0.5cm}}

 \author{Yiming Chen${}^{1,2}$ and Gustavo J. Turiaci${}^{3,4}$}

\affiliation{${}^1$ Stanford Institute for Theoretical Physics, Stanford, CA USA}
\affiliation{${}^2$ Physics Department, Princeton University, Princeton, NJ USA}
\affiliation{${}^3$ Physics Department, University of Washington, Seattle, WA USA}
\affiliation{${}^4$ Institute for Advanced Study, Princeton, NJ USA}

\abstract{ The gravitational path integral can be used to compute the number of black hole states for a given energy window, or the free energy in a thermal ensemble. In this article we explain how to use the gravitational path integral to compute the separate number of bosonic and fermionic black hole microstates. We do this by comparing the partition function with and without the insertion of $(-1)^{\sf F}$. In particular we introduce a universal rotating black hole that contributes to the partition function in the presence of $(-1)^{\sf F}$. We study this problem for black holes in asymptotically flat space and in AdS, putting constraints on the high energy spectrum of holographic CFTs (not necessarily supersymmetric). Finally, we analyze wormhole contributions to related quantities.   }

\begin{document}\maketitle

\section{Introduction} 

The gravitational path integral \cite{Gibbons:1976ue} is normally used to compute the total number of black hole microstates with some constraints such as a fixed energy window or temperature. The result is usually phrased in terms of the free energy $F(\beta)$ as a function of temperature or the microcanonical entropy $S(E)$ as a function of the energy
\bea\label{eq:pf}
Z(\beta) = {\rm Tr} \,\, e^{-\beta H } = e^{- \beta F(\beta)},~~~~{\rm Tr}_E \,\,1 = e^{S(E)}.
\ea
The purpose of this article is to apply this formalism to derive statistical properties of how these black hole microstates are distributed according to whether they are bosonic or fermionic. In principle, for Lorentz invariant theories the quantum statistics of a state is correlated with the angular momentum: states with half-integer angular momentum are fermions while states with integer angular momentum are bosons. This would correctly suggest we can estimate the quantum statistics of a black hole microstate by looking at its angular momentum. The problem with this approach is that in the limit of small $G_N$, where we can rely on semiclassical gravity, charges including the angular momentum are large and of order $1/G_N$. The distinction of whether the angular momentum is integer or half-integer is therefore beyond the classical approximation. The determination of the quantum statistics from the gravitational path integral in the way described in this paragraph therefore requires incorporating quantum effects. 

To extract the distribution of fermionic and bosonic black hole microstates in a simpler way, we focus on the following quantities which generalize \eqref{eq:pf}:
\beq\label{eq:pfspin}
Z_\spin(\beta) = {\rm Tr} \, (-1)^{\sf F} e^{-\beta H} = e^{- \beta F_\spin(\beta)},~~~~~{\rm Tr}_E \,(-1)^{\sf F} = e^{S_\spin (E)}.
\eeq
These are the free energy $F_\spin(\beta)$ as a function of temperature, and the entropy $S_\spin(E)$ as a function of energy, both computed with an insertion of $(-1)^{\sf F}$ the operator that assigns a value of $+1$ to bosonic states and $-1$ to fermionic states. The quantities in \eqref{eq:pfspin} together with \eqref{eq:pf} are enough to extract the number of bosonic against fermionic black hole microstates. The reason to study \eqref{eq:pfspin} is that, as we explain in this paper, this quantities can be computed using the gravitational path integral in the classical approximation. 

The idea to evaluate \eqref{eq:pfspin} is the following. For concreteness we frame the discussion in the context of ${\rm AdS}_{d+1}/{\rm CFT}_d$ and use the gravitational path integral to compute the partition function of the boundary CFT on $S^1 \times S^{d-1}$ both with and without an insertion $(-1)^{\sf F}$. Let us consider the fixed temperature ensemble since similar considerations apply when fixing energy. When evaluating the gravitational path integral, since the thermal partition function $Z(\beta)$ is an Euclidean observable it gets a contribution from the Euclidean section of a non-rotating black hole as well as from thermal AdS. In principle the only difference between the partition function with or without an insertion of $(-1)^{\sf F}$ is whether fermionic fields are periodic or anti-periodic around the time cycle, so the same black hole and thermal AdS saddles should contribute to both quantities. Nevertheless, in the presence of the $(-1)^{\sf F}$ insertion, fluctuations of fermionic fields around the black hole saddle are singular since the choice of spin structure is incompatible with the time circle contracting at the horizon. 

So what saddles do contribute to $F_\spin$ or $S_\spin$? The first obvious option is thermal AdS. Since the time circle does not contract anywhere inside the thermal AdS geometry we can evaluate $F_\spin$ or $S_\spin$ by counting bosonic vs fermionic low energy exitations around vacuum AdS. To leading order in the small $G_N$ limit this predicts a small value for $F_\spin, S_\spin \sim \mathcal{O}(1)$. This is much smaller than the results without the $(-1)^{\sf F}$ insertion since $F, S \sim \mathcal{O}(1/G_N)$, dominated by the black hole saddle. If this was true, it implies most states are evenly distributed between bosons and fermions and there is a large cancellation in $F_\spin$ or $S_\spin$. Some evidence for a huge cancellation between bosonic and fermionic states in a theory without supersymmetry was presented from the boundary side in \cite{Cherman:2018mya} (and references therein).  

\smallskip 

The purpose of this article is to introduce another universal saddle that always contributes to both $F_\spin$ and $S_\spin$, analogous to the black hole contribution to $F$ or $S$. For concreteness we focus first on the free energy. Choose an arbitrary direction and define $J$ as the angular momentum generating rotations around this direction. One can generalize \eqref{eq:pf} and compute $ {\rm Tr}\,\left( e^{-\beta H} e^{\beta \Omega J} \right)$, where $\Omega$ is the angular velocity of the ensemble. On the boundary side this is implemented by imposing a twist in the boundary conditions: as fields go around the time cycle they get multiplied by $+e^{\beta \Omega J}$ (bosons) or $-e^{\beta \Omega J}$ (fermions). The black hole that contributes to this quantity in the gravitational path integral has now rotation. This implies that the contractible cycle is not time anymore, but a combination of time and the angle around the $J$ direction.  This partition function becomes $F_\spin$ when $\beta \Omega = 2\pi \i$, since $(-1)^{\sf F} = e^{ 2 \pi \i J}$. Therefore there is a universal rotating black hole saddle that contributes to $F_\spin$ such that periodic fermions around the time cycle produces a smooth spin structure. 

The way described in the previous paragraph of generating black hole solutions with periodic fermions was considered exclusively for supersymmetric solutions in \cite{Cabo-Bizet:2018ehj, Bobev:2020pjk, Iliesiu:2021are}. The point of the present paper is to apply this same construction to general theories that are either not supersymmetric, or supersymmetric but the partition function with a $(-1)^{\sf F}$ insertion is not protected. 

\smallskip 

In the next sections we study the rotating black hole geometries that contribute to $F_\spin$ and $S_\spin$ and analyze their consequences. For the spacetime dimensions we analyzed, we always find that the rotating black hole saddle that contribute to $F_\spin$ has a higher free energy than thermal AdS with periodic fermions making it always subleading. The free energy $F$ has a phase transition as a function of temperature when thermal AdS and the black hole change dominance, which on the boundary side is interpreted as a confinement-deconfinement phase transition \cite{Witten:1998zw}. Instead, even after including the rotating black hole saddle, the free energy $F_\spin$ in the presence of $(-1)^{\sf F}$ has no phase transition and depends smoothly on the temperature being always controlled by thermal AdS. In the presence of bulk gauge fields, it is easy to remedy this and make our rotating black hole dominant: one can consider an ensemble where some charge is fixed such that thermal AdS is discarded.  

A different way to make the rotating black hole be dominant is to work in the fixed energy ensemble and compute the entropy $S_\spin$. The rotating black hole does contribute a positive and large amount to $S_\spin$. For illustration take ${\rm AdS}_4$. The entropy without and with the insertion of $(-1)^{\sf F}$ are given, at large energies, by (see Section \ref{sec:ads4} for conventions) 
\beq
S(E) \sim E^{2/3} ,~~~~~S_\spin (E) \sim E^{1/2}.
\eeq
The black hole result $S(E) \sim E^{2/3}$ is consistent with a local 3d theory on the boundary. The rotating black hole contributing to $S_\spin$ gives a positive contribution that grows with energy, albeit slower than the total number of states. The results for ${\rm AdS}_3$ are similar qualitatively. Instead, in ${\rm AdS}_5$ we find that both $F_\spin$ and $S_\spin$ are dominated by thermal AdS.

We also study the contribution from wormholes to the partition function in the presence of a $(-1)^{\sf F}$ insertion. More precisely, we study a generalization of the spectral form factor \cite{Cotler:2016fpe,Saad:2018bqo} where $(-1)^{\sf F}$ is inserted. We discuss a construction similar to the double cone of \cite{Saad:2018bqo} and evaluate its contribution to the gravitational path integral. We argue that at late time the spectral form factor is not sensitive to the $(-1)^{\sf F}$ insertion.

\smallskip 

We finish this introduction with some general comments and relation to other work. A universal formula was derived in \cite{Pal:2020wwd,Harlow:2021trr} for the density of states for theories with finite-group symmetry. If the finite-group is $\mathbb{Z}_2$ the formula predicts that high energy states are equally distributed between even and odd states. In our context the finite-group is generated by $(-1)^{\sf F}$ and their result would apply if no saddle would contribute an amount to $F_\spin$ (or $S_\spin$) comparable to $F$ (or $S$), such that to leading order there are as many bosonic as fermionic states. In this context, our rotating black hole provides a universal geometric configuration that gives a subleading correction to the result of \cite{Pal:2020wwd,Harlow:2021trr}. Of course there can be other contributions such as defects which provide other corrections to \cite{Pal:2020wwd,Harlow:2021trr} (see the discussion section for further comments), but such a contribution would not be universal and would depend on the details of the defect and the theory in which they are embedded.  

Another interesting feature of our rotating black hole solution is that it does not respect ensemble equivalence: the saddle that dominates at fixed temperature (thermal AdS) is completely different than the saddle that dominates the microcanonical ensemble (the rotating black hole). In Section \ref{sec:AdS} we give an interpretation of why this is so. In the context of ${\rm AdS}_3/{\rm CFT}_2$ Tauberian theorems that incorporate rotation \cite{Mukhametzhanov:2020swe,Pal:2022vqc, Pal:2023cgk} might be powerful enough to prove when different ensembles are equivalent or not and therefore studying $Z_\spin$ in that context could be interesting.  

\bigskip 

The rest of the paper is organized as follows. In Section \ref{sec:bh} we find new black hole geometries that contribute to the partition function with an insertion of $(-1)^{\sf F}$. In Section \ref{sec:4dflat} we study this problem for black holes in flat space, with and without charge, and in Section \ref{sec:AdS} we generalize this to ${\rm AdS}_4$, ${\rm AdS}_5$ and finally ${\rm AdS}_3$.  In Section \ref{sec:wormhole} we study the contribution from wormholes to the partition function in the presence of a $(-1)^{\sf F}$ insertion. We conclude in Section \ref{sec:discussion} with discussions of our results and future directions, leaving technical details for appendices.

\section{Black hole solutions in the presence of $(-1)^{\sf F}$}\label{sec:bh} 

\subsection{Flat Space Solutions} \label{sec:4dflat}
In this section we explain the general principle to compute the distribution of fermionic and bosonic black hole microstates using the gravitational path integral. We begin by studying black holes in asymptotically four dimensional flat spacetime. As we will see, the calculation in flat space is a bit singular. However we decide to include it since it is a good starting point to illustrate the main ideas. The action, in Euclidean signature, is given by 
\beq\label{eq:EHaction}
I = -\frac{1}{16\pi G_N} \int \sqrt{g} R -\frac{1}{8\pi G_N}\oint \sqrt{h} K + ({\rm matter}).
\eeq
The details of the matter sector will not be important other than the assumption that part of it is fermionic. This assumption does not show up explicitly in the following discussion, while its importance will be discussed in Section \ref{sec:discussion}. This theory has classical black hole solutions described by the Kerr metric, which in Boyer-Lindquist coordinates is given by
\beq\label{eq:Kerr}
\d s^2 = - \frac{\rho^2 \Delta}{\Xi} \d t^2 + \frac{\rho^2}{\Delta} \d r^2 + \rho^2 \d \theta^2 + \frac{\Xi}{\rho^2} \sin^2 \theta\left( \d \varphi - \frac{2Ear}{\Xi} \d t \right)^2 
\eeq
with the functions
\bea \label{eq:Kerrdef}
&& \rho^2 = r^2 + a^2 \cos^2 \theta, \nonumber\\
&& \Delta= r^2 - 2 E r + a^2, \\
&& \Xi = (r^2 + a^2)^2 - a^2 \Delta \,\sin^2 \theta . \nonumber
\ea
$E$ and $a$ are parameters of the solution, and their interpretation depends on the ensemble we choose to work with. In a microcanonical ensemble, $E$ is the ADM mass while $J = a E$ is the angular momentum. In the classical limit both of these quantities are large, of order $1/G_N$ and one is \emph{a priori} not sensitive to questions such as whether $J$ is  integer (bosonic) or half-integer (fermionic). Below we work in units such that $G_N=1$. 

In the grandcanonical ensemble, we want instead to fix the inverse temperature $\beta$, the length of the thermal circle, and the angular velocity $\Omega$ at infinity. For the metric given in equation \eqref{eq:Kerr} this is achieved by imposing the following identification
\beq\label{eq:identGCFS}
(t_{\rm E}, \varphi) \sim (t_{\rm E}+\beta,\varphi+\i \beta \Omega) \sim (t_{\rm E},\varphi + 2\pi),
\eeq
where $t_{\rm E} = - \i t$ is the Euclidean time.  In this ensemble, the parameters $a$ and $E$ should be seen as being fixed by the temperature and angular velocity in the following way. The Kerr metric has an outer event horizon at $r_+ = E + \sqrt{E^2 - a^2}$. It will be useful sometimes below to trade the parameter $E$ by $r_+$ since $E=\frac{r_+^2 + a^2}{2r_+}$. Demanding that the solution \eqref{eq:Kerr} is smooth at the horizon, given the identification \eqref{eq:identGCFS}, determines the parameters $a$ and $E$ (or $r_+$) as functions of $\beta$ and $\Omega$ by
\beq\label{eq:Pote}
\beta = \frac{4\pi r_+ (a^2 + r_+^2)}{(r_+^2-a^2)},~~~~\Omega = \frac{a}{r_+^2 + a^2} .
\eeq
It is for this choice of ensemble that the action quoted in equation \eqref{eq:EHaction} with the Gibbons-Hawking-York boundary term poses a well-defined variational problem. 

\smallskip 

Let us begin by computing the partition function in the grandcanonical ensemble with $\Omega=0$, where we sum over states with arbitrary mass and angular momentum. This is of course a very well known result \cite{Gibbons:1976ue}. Classically, $\Omega=0$ implies that $a=0$ and then the metric \eqref{eq:Kerr} reduces to the Schwarszchild black hole. For any value of $\beta$ and $\Omega$, the action \eqref{eq:EHaction} evaluated on this solution is equal to $I =\beta E- A/4 - \beta \Omega J$, where $A = 4\pi (r_+^2 + a^2)$ is the area of the event horizon. For $\Omega=0$ the free energy then is
\beq
- \beta F = \log Z =  - \frac{\beta^2}{16 \pi },~~~~F= \frac{\beta}{16\pi}.
\eeq
In a quantum mechanical description of the black hole microstates, this quantity corresponds to $Z(\beta) = {\rm Tr} \, e^{-\beta H}$, where the trace is taken over states of any mass or angular momentum. We can also work in an ensemble of fixed mass, and again restrict to $\Omega=0$. We refer to this partition function as $e^{S(E)}$, and can be interpreted as the number of states of energy $E$ regardless of the spin.\footnote{In the semiclassical approximation it makes more sense to specify the energy $E$ to be in a small window $(E-\delta E, E + \delta E)$ with small enough $\delta E$. Whenever we specify the energy from now on we always mean up to $\delta E$.} On the black hole side the energy corresponds to the mass $E$. This quantity can be obtained either by Legendre transform of the result at finite temperature, or by finding the appropriate boundary terms in the action. Either way the answer is 
\beq
S(E) = 4\pi  E^2.
\eeq
This is the standard Bekenstein-Hawking area law for a Schwarszchild black hole.

The result above gives a prediction for the total number of black hole microstates of a given energy, ${\rm exp}\{ S(E)\}$. The question we want to address here is what fraction of these states are fermionic states and what fraction are bosonic. On general grounds, we expect that both bosonic and fermionic microstates constitute about half of the total number of states. This is because we can turn a bosonic black hole into a fermionic one by throwing in a fermion, which barely changes its energy. However, we do not expect the ratio to be precisely one half. We might have slightly more bosonic states or fermionic states depending on the energy window we are looking at. It is this subtle difference that we are after. 

\smallskip 

To approach this question, we analyze a quantity whose evaluation is easy to formulate using the gravitational path integral. In terms of the quantum system describing the black hole microstates we would like to compute the partition function with a fermion parity operator, an insertion of $(-1)^{\sf F}$.\footnote{As we mentioned in the introduction, the quantity $\textrm{Tr} \, (-1)^{\sf F} e^{-\beta H}$ is commonly studied for supersymmetric theories, and is called the Witten index \cite{Witten:1982df}. Here we are \emph{not} considering supersymmetric theories.} This is 
\beq
{\rm Tr} \, (-1)^{\sf F}e^{-\beta H} = e^{- \beta F_{\spin}},~~~~~{\rm Tr}_E \,(-1)^{\sf F} = e^{S_{\spin}(E)}
\eeq
in an ensemble with fixed temperature or energy, respectively. We will use the gravitational path integral to compute the free energy $F_\spin$ and entropy $S_\spin$ in the presence of a $(-1)^{\sf F}$ insertion, to leading order in the classical limit. 

\begin{figure}
    \centering
  \begin{tikzpicture}
\draw[black!80, thick,->] (0,0) -- (4,0);
\draw[black!80, thick,->] (0.2,-0.2) -- (0.2,4.2);
\node at (-0.05,3) {$\beta$};
\node at (-0.15,4.2) {$t_{\rm E}$};
\node at (2.2,-0.3) {$2\pi$};
\node at (4,-0.25) {$\varphi$};
\draw[blue!70, ultra thick,->] (0.2,0) -- (1.5,3);
\draw[red!70, ultra thick,->] (0.2,0) -- (0.2,3);
\draw[gray, dashed, thick] (0.2,0) -- (1.5,3) -- (1.5+2,3) -- (0.2+2,0);
\draw [decorate,decoration={brace,amplitude=10pt}] (0.2,3) -- (1.5,3) node [black,midway,yshift=17pt] {$\i \beta \Omega$};
\end{tikzpicture}
\hspace{3cm}
 \begin{tikzpicture}
\draw[black!80, thick,->] (0,0) -- (4,0);
\draw[black!80, thick,->] (0.2,-0.2) -- (0.2,4.2);
\node at (-0.05,3) {$\beta$};
\node at (-0.15,4.2) {$t_{\rm E}$};
\node at (2.2,-0.3) {$2\pi$};
\node at (4,-0.25) {$\varphi$};
\draw[blue!70, ultra thick,->] (0.2,0) -- (2.2,3);
\draw[red!70, ultra thick,->] (0.2,0) -- (0.2,3);
\draw[gray, dashed, thick] (0.2,0) -- (0.2+2,3) -- (0.2+2+2,3) -- (0.2+2,0);
\draw [decorate,decoration={brace,amplitude=10pt}] (0.2,3) -- (2.2,3) node [black,midway,yshift=17pt] {$\i \beta \Omega=2\pi$};
\end{tikzpicture}
    \caption{\footnotesize $(t_{\rm E}, \varphi)$-plane at fixed $r$ and $\theta \neq 0, \pi$. Working in the grandcanonical ensemble implies two identifications given in equation \eqref{eq:identGCFS}, the fundamental domain is bounded by the dashed lines. We show the thermal cycle $(t_E, \varphi) \sim (t_E + \beta , \varphi)$ (red arrow) and the cycle that contracts at the horizon (blue arrow). \textit{Left:} General value of $\Omega$. Fermions are always antiperiodic around the blue cycle and around $\varphi \sim \varphi + 2\pi$ since they are both contractible in the black hole geometry. Fields get multiplied by $e^{\beta \Omega J}$ (bosons) or $-e^{\beta \Omega J}$ (fermions) as they go around the red cycle. \textit{Right:} Choice of angular velocity that computes the partition function with a $(-1)^{\sf F}$ insertion. $\Omega$ is such that fermions are antiperiodic along the blue cycle but now become periodic around the red one since $e^{2\pi \i J} = 1$ (bosons) or $-1$ (fermions).}
    \label{fig:ident}
\end{figure}
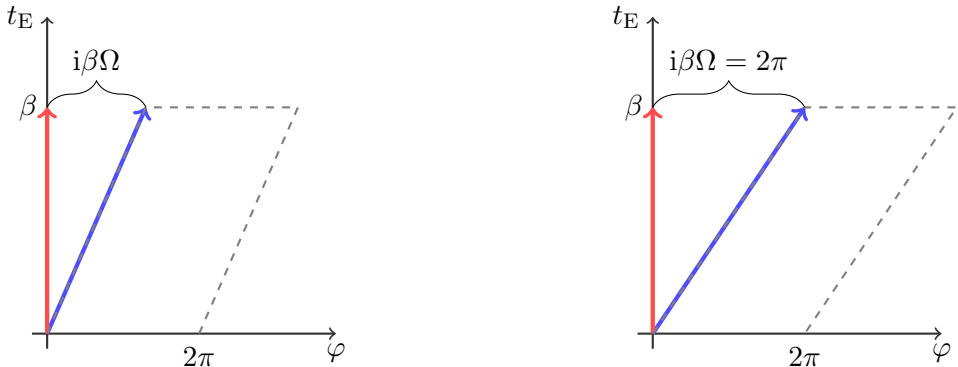

In a naive approach to this question, one would conclude that with the insertion of $(-1)^{\sf F}$ we should study the same Schwarszchild black hole geometry, but impose that fermions should be periodic around the thermal circle. Since this circle contracts at the horizon, this is not a smooth choice of spin structure. This approach is naive since we can generate periodic boundary conditions for fermions starting from \eqref{eq:identGCFS} and setting instead $\beta \Omega = \pm 2\pi \i$. Now the solution is the Kerr black hole and the rotation implies the time cycle does not contract at the horizon anymore. Instead it is precisely the one along the first identification in equation \eqref{eq:identGCFS} that contracts and fermions are antiperiodic around this cycle \cite{Iliesiu:2021are}. We illustrate this argument  in figure \ref{fig:ident}. Another way to think about this solution is to write 
\beq\label{eq:relFJ}
(-1)^{\sf F} = e^{\pm 2\pi \i J}.
\eeq
Using this relation, the insertion of $(-1)^{\sf F}$ corresponds to a specific choice of the angular velocity. 

Let us make a few clarifications. The boundary conditions corresponding to the grandcanonical ensemble given in \eqref{eq:identGCFS} break the full rotational isometries of flat space down to the one generated by $\partial_\varphi$. This is true for generic values of $\Omega$. An exception is when $\Omega=0$: the full $SU(2)$ rotational symmetry is restored in the boundary and the bulk geometry, the Schwarszchild metric, preserves these isometries everywhere in the interior of the spacetime. Another exception is when $\beta \Omega = \pm 2\pi \i$: the relation \eqref{eq:relFJ} is independent of the choice of axis for $J$ in the right hand side and therefore the boundary conditions preserve the full rotational invariance. Nevertheless, the new feature is that now the bulk geometry, the Kerr metric, does break the rotational isometries down to $\partial_\varphi$ since it requires singling out an axis of rotation. This implies the presence of a moduli space of Kerr black holes with arbitrary axis of rotation, and the rotational invariance arises from integrating over this moduli space. The presence of these continuum bosonic zero-modes is not problematic since the moduli space is compact\footnote{This was pointed out in \cite{Iliesiu:2021are}
and analyzed in detail in \cite{H:2023qko} in the context of concrete supergravity examples.}, and in any case this issue would only arise when evaluating quantum corrections which we will not attempt here. Given the choice of axis we made here, the two values $\beta \Omega = + 2\pi \i$ or $- 2\pi  \i$ are connected inside the moduli space, and therefore we can restrict from now on to $\beta \Omega = + 2\pi \i$ for concreteness. 

After these generalities, let us  compute $F_\spin$ using the rotating black hole solution. We first need to impose $ \beta \Omega = 2\pi \i $ and solve for $r_+$, or equivalently $E$, and $a$. More generally, we have the following relation between $r_+$ and $a=\i {\sf a}$ 
\beq
\frac{2 {\sf a} r_+}{{\sf a}^2 + r_+^2} = \frac{\beta \Omega}{2\pi \i} \equiv \omega.
\eeq
We want to set $\omega = 1$. Assuming $r_+$ is finite, this leads to ${\sf a} = \pm r_+$ which in turn implies $\beta=0$. This is an issue since in principle we should be able to compute the free energy for any temperature using the gravitational path integral. A way out is to consider the fixed energy ensemble instead, which we do below. Another option is to work with $\omega = 1 - \varepsilon$ and take $\varepsilon \to 0$ at the end of the calculation. After eliminating $a$, the inverse temperature is given by 
\beq
\beta = 4 \pi r_+ \sqrt{\varepsilon (2-\varepsilon)},~~\Rightarrow~~r_+ = \frac{\beta}{4\pi \sqrt{\varepsilon (2-\varepsilon)}}.
\eeq
This expression is valid for any $\varepsilon$, but in the $\varepsilon \to 0 $ limit  we get $r_+ \sim 1/\sqrt{\varepsilon} \to \infty$. It might be worrisome that computing $F_\spin$ involves a geometry with $r_+ \to \infty$. Nevertheless one can check that in this limit the energy, the regularized action, and the curvature tensor, all remain finite as explained in Appendix \ref{app:KerrR2}. (This issue will not arise for black holes in AdS.) Thus we will accept this solution and continue our analysis. We can estimate the contribution of this saddle to the difference between fermionic and bosonic states by computing the on-shell action. We obtain
\beq
I = \beta E- \frac{A}{4} - \beta \Omega J = \frac{\beta^2}{8 \pi (1 + \sqrt{\varepsilon(2-\varepsilon)})} ,
\eeq
for an arbitrary value of $\varepsilon$. Taking the limit $\varepsilon \to 0$ gives $I = \beta^2 / 8\pi$ and therefore 
\beq
-\beta F_\spin = - \frac{\beta^2}{8\pi},~~~~F_\spin = \frac{\beta}{8\pi}. 
\eeq
To interpret this result, we repeat this analysis in the fixed energy ensemble. In this case, instead of determining $r_+$ from $\beta$, $r_+$ is determined by the fixed mass $E$. We obtain $r_+ \sim E/\sqrt{4\varepsilon}$ in the small $\varepsilon$ limit. In the fixed energy ensemble the classical action has an extra boundary term $-\beta E$ and the total answer for the action is
\beq
I_{\rm fixed~E} = - \frac{A}{4} - \beta \Omega J \to 2 \pi E^2.
\eeq
Therefore the entropy in the presence of $(-1)^{\sf F}$, counting the difference between fermionic and bosonic black hole microstates of mass $E$ in the semiclassical limit, is given by 
\beq
S_\spin(E)=2 \pi E^2 = \frac{1}{2} S(E).
\eeq 
This implies that if the total number of black hole microstates of energy $E$ grows as $N_{\rm tot}$, then the  absolute value of the difference in number of bosonic minus fermionic states grows as $\sqrt{N_{\rm tot}}$. Notice at the classical level we cannot determine which statistic has more states. To determine the overall sign would require evaluating the one-loop determinant.

A possible interpretation for the singular behavior of the flat space calculation could be the following. The flat space black hole is coupled to radiation extending to infinity. In flat space, we have the possibility of introducing very light particles far away from the black hole that carry large angular momentum. This suggests that in this case it might be subtle to define a closed system involving the black hole and its environment, for which the trace $\textrm{Tr}\,(-1)^{\sf F}$ can be defined. Such a problem does not exist in AdS space, and as we will see in Section \ref{sec:AdS} the calculation is indeed non-singular. 

\bigskip 

There is a final point we want to address before moving on. Since we assume that the theory includes fermionic and bosonic fields, a background can be probed by operators of half-integer or integer angular momentum. Therefore the boundary condition in equation \eqref{eq:identGCFS}  only depends on $\i \beta \Omega ~{\rm mod}~4\pi \mathbb{Z}$, while the black hole solution depends on its real valued lift $\i \beta \Omega$. This is resolved by summing over saddles obtained by integer shifts $\i\beta \Omega \to \i \beta \Omega + 4 \pi \mathbb{Z}$. What subset of these saddles should be included is an open question, and affects the entropy both with and without the insertion of $(-1)^{\sf F}$. We leave this for future work.

\subsubsection*{Charged black hole}

We can extend the previous calculation to black holes in the presence of a Maxwell field, in a background of fixed electric charge $Q$, which we take to be positive. The classical solution of Einstein-Maxwell theory we would need in this case is the Kerr-Newman black hole. We normalize the charge $Q$ such that the thermodynamic potentials derived from the Kerr-Newman solution are
\beq\label{eq:betaOmegaKN}
 \beta = \frac{4 \pi r_+ (r_+^2 + a^2)}{r_+^2 - a^2 -Q^2} ,~~~\Omega = \frac{a}{r_+^2 + a^2} 
\eeq
while the mass is given in terms of $r_+$ as $E=\frac{r_+^2 + a^2 + Q^2}{2r_+}$ and $J= aE$. The saddle contributing to the grandcanonical partition function, with $\Omega=0$ and therefore no $(-1)^{\sf F}$ insertion, has $a=0$. This leads to the Euclidean section of the Reissner-Nordstrom black hole. At fixed temperature, the on-shell action has an explicit but complicated expression at fixed temperature. The result simplifies when written in the fixed energy ensemble and results in an entropy
\beq
S(E,Q)= \pi \big(E+\sqrt{E^2 - Q^2}\big)^2,
\eeq
the standard Bekenstein-Hawking area law for the Reissner-Nordstrom black hole. We only consider the black hole solutions with $E\geq Q$. 

We can use the gravitational path integral to compute the difference between the number of bosonic and fermionic black hole microstates. To do this, for the same reasons as outlined before, we need to set $\beta \Omega = 2\pi \i$. This implies, using equation \eqref{eq:betaOmegaKN} and writing $a=\i {\sf a}$, that $2 {\sf a} r_+ = r_+^2 +{\sf a}^2 - Q^2$ and therefore ${\sf a}= r_+ - Q$. The horizon radius is fixed in terms of the inverse temperature as $r_+ = \frac{Q(\beta - 2 \pi Q)}{\beta-4\pi Q}.$\footnote{The radius of the horizon is negative for $2\pi Q <\beta < 4\pi Q$. One option to deal with temperatures in this range is to define the geometry along a complex contour in the $r$-plane that approaches $r_+$ from infinity without going close to the singularity at $r=0$. This type of complex geometry would violate the criterion put forth in \cite{Witten:2021nzp}. For $\beta< 2 \pi Q$, $r_+$ is positive again but a new horizon appears with a larger value of $r$, which again has to be avoided along the complex plane. If we consider the classical limit of $\mathcal{N}=2$ supergravity in 4d, these complex geometries seem necessary to ensure that the index is temperature independent.} In this case the on-shell action appropriate to the fixed temperature ensemble has a very simple form $I= -\pi Q^2 + \beta Q$, implying that 
\beq
- \beta F_\spin = \pi Q^2 - \beta Q, 
\eeq
which is the result found in \cite{Iliesiu:2021are}. 

This solution has two issues. The first is that it does not have a smooth $Q \to 0$ limit since $r_+ \to 0$. The second is that for this solution the energy is $E=Q$ independently of $\beta$. Instead we should be able to compute the difference between bosonic and fermionic microstates in a fixed energy ensemble with any $E\neq Q$. This requires finding a different saddle in the fixed temperature ensemble. To do this set $\beta \Omega = 2\pi \i (1-\varepsilon)$ and take the limit $\varepsilon \to 0$ more carefully. This way we discover a new solution such that the horizon radius is 
\beq\label{eq:rpqfos}
r_+ \sim \frac{\sqrt{\beta^2 -(4\pi Q)^2}}{4\pi\sqrt{2\varepsilon}} ,~~~\varepsilon \ll 1.
\eeq
This solution is similar to the one we found in the uncharged case. Now the average ADM energy is given by $E = \beta / 4\pi$ and becomes a free parameter set by the temperature (and happens to be independent of the charge). The action in the fixed temperature ensemble is now $I = \pi Q^2 + \beta^2 / 8\pi$, or equivalently
\beq
-\beta F'_\spin = - \pi Q^2  - \frac{\beta^2}{8\pi}.
\eeq
Notice the sign change of the temperature independent term between $F_\spin$ and $F_\spin'$. The true free energy should be the minimum between $F_\spin$ and $F'_\spin$, which happens to be $F_\spin$ for any temperature.

The goal of \cite{Iliesiu:2021are} was to use this type of solution to compute the index of the black hole in $\mathcal{N}=2$ supergravity. In this case we expect the solution with $E=Q$ to be the only contribution. The boundary conditions in this context are supersymmetric and any solution that breaks supersymmetry (if $E\neq Q$ there is no Killing spinor) will have a fermionic zero-mode that makes this contribution vanish.\footnote{Even if $E=Q$ some supersymmetries are broken since the Reissner-Nordstrom preserves only four supercharges, and the contribution to the index also vanishes. One can fix this considering the helicity supertrace instead of the index, which can be also computed with the gravitational path integral.} The second solution above with $r_+ \to \infty$ therefore makes a vanishing contribution to the index in supergravity due to this fermion zero-modes. If we are working with a non-supersymmetric theory of gravity, then both solutions should be in principle included. For any temperature, the black hole with $r_+ $ finite has always a lower free energy  and is dominant. Nevertheless, when working at fixed energy $E>Q$ the solution with finite $r_+$ disappears and only the solution with $r_+ \to \infty$ remains.
\begin{figure}
\begin{center}
\includegraphics[scale=0.7]{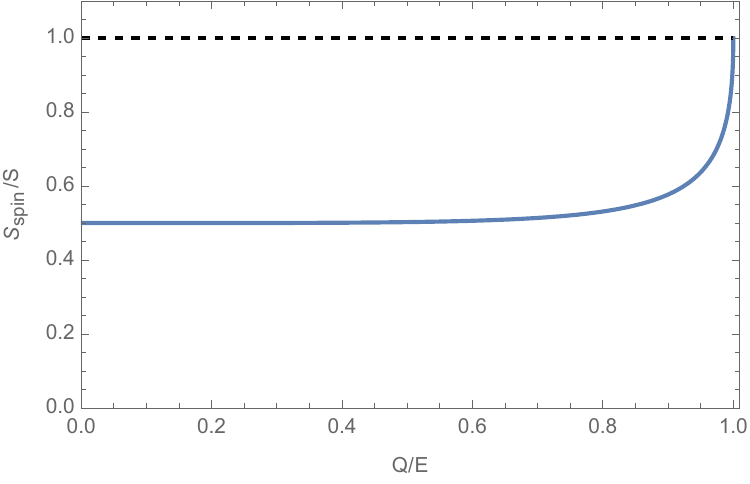}
\end{center}
\vspace{-5mm}
\caption{\footnotesize Plot of $S_\spin/S$ for a charged black hole of energy $E$ as a function of $q=Q/E$. It is always smaller than one and for a given energy $E$, it interpolates between $1/2$ for $Q=0$ and $1$ for $Q=E$.}
\label{fig:4dflat}
\end{figure}

Let us then consider entropy in the fixed energy ensemble with the insertion of $(-1)^{\sf F}$, using the relation between energy and temperature derived above, $\beta = 4 \pi E$. The action, including the boundary terms necessary to change ensemble, is given by $\pi (2E^2 - Q^2)$ and therefore the entropy counting the difference between bosonic and fermionic microstates is 
\beq
S_\spin(E,Q)= \pi (2E^2 -Q^2).
\eeq
We do not expect any black hole microstate with $E<Q$  regardless of its statistics, so this result should only be considered for $E>Q$. The ultimate reason why a complex saddle with $E<Q$ is not allowed has to be tracked back to the integration contour of the gravitational path integral. 

In the uncharged case we found $S_\spin (E,Q=0) = \frac{1}{2} \cdot S(E,Q=0)$. We can now evaluate the ratio between the entropy with and without a $(-1)^{\sf F}$ insertion and we find the following charge dependent result
\beq
\frac{S_\spin(E,Q)}{S(E,Q)} = \frac{2-q^2}{(1+\sqrt{1-q^2})^2}\leq 1,~~~q\equiv \frac{Q}{E},
\eeq
where $0\leq q \leq 1$. We plot this function in figure \ref{fig:4dflat}. When $q=0$ we recover the $1/2$ we found for the uncharged black hole. When $q=1$ we reproduce the expectation that number of states counted with a $(-1)^{\sf F}$ is the same as the total number of states, and therefore to leading order all states have the same statistics. (This interpretation is not correct after including quantum effects which imply the zero-temperature entropy at $E=Q$ does not correspond to a real degeneracy \cite{Ghosh:2019rcj,Iliesiu:2020qvm,Heydeman:2020hhw,Turiaci:2023wrh}.) As indicated above, this ratio is always smaller than one. This is consistent with unitarity since the total number of states cannot be smaller than the number of bosonic minus fermionic states.

\subsection{AdS Solutions}\label{sec:AdS}
\subsubsection{AdS${}_4$}\label{sec:ads4}
In this section we analyze the quantum statistics of black hole microstates in asymptotically AdS${}_4$ spaces. Through AdS/CFT this puts a constraint on the quantum statistics of high energy states of holographic CFT${}_3$'s.  

As explained for example in \cite{Chamblin:1999tk} it is possible to construct backgrounds in string theory where the low energy theory is described by Einstein-Maxwell in asymptotically AdS${}_4$ coupled to matter
\beq
I = -\frac{1}{16 \pi } \int \d^4 x \sqrt{g} [ R - 2 \Lambda - F^2] + I_{\rm bdy}.
\eeq
We work in units with $G_N=1$. We parametrize the cosmological constant by $\Lambda = -3/\ell^2$ and $I_{\rm bdy}$ are the boundary terms necessary depending on the ensemble. The black hole solutions are given by the following generalization of the Kerr-Newman black hole
\beq \label{eq:ads4}
\d s^2 = -\frac{\Delta_r}{\rho^2} \Big[ \d t - \frac{a \sin^2\theta}{\Xi} \d \phi \Big]^2 + \frac{\rho^2}{\Delta_r} \d r^2 + \frac{\rho^2}{\Delta_\theta} \d \theta^2 + \frac{\Delta_\theta \sin^2 \theta}{\rho^2} \left[ a \d t - \frac{r^2 + a^2}{\Xi}\d \phi \right]^2
\eeq
where 
\bea
&&\rho^2 = r^2 + a^2 \cos^2 \theta,~~~\Xi = 1- a^2/\ell^2,\\
&&\Delta_r = (r^2+a^2)(1+\frac{r^2}{\ell^2}) - 2 m r + q^2 ,~~~\Delta_\theta = 1-\frac{a^2}{\ell^2 } \cos^2 \theta.
\ea
We take the charge to be electric and the gauge field is $A = -\frac{q r}{\rho^2}(\d t - \frac{a \sin^2 \theta}{\Xi} \d \phi)+\alpha \d t$, where $\alpha$ is a constant chosen to make the gauge connection smooth at the horizon. The solution is parametrized by $m$, $a$, and $q$, which roughly correspond to the mass/energy (or temperature), the angular momentum (or angular velocity), and the charge (or chemical potential). 

In the Euclidean section, one can find the inverse temperature and angular velocity by demanding smoothness at the horizon, located at $\Delta_r (r_+)=0$, which gives 
\beq
\beta = \frac{4\pi (r_+^2 + a^2)}{r_+(1+\frac{a^2}{\ell^2} + 3\frac{r_+^2}{\ell^2} - \frac{a^2 +q^2}{r_+^2})},~~~\Omega = \frac{a (1+r_+^2/\ell^2)}{r_+^2 + a^2}.
\eeq
We can trade the parameter $m$ by $r_+$ using the relation $m = \frac{(r_+^2 + a^2)(1+\frac{r_+^2}{\ell^2})+q^2}{2r_+}$. The ADM charges, the energy $E$, angular momentum $J$, and $\U(1)$ charge $Q$ are given by
\beq
E = \frac{m}{\Xi^2},~~~~J=a E,~~~~Q=\frac{q}{\Xi}.
\eeq
Finally, the area of the even horizon is given by $A= 4\pi (r_+^2 + a^2)/\Xi$.  

Below we consider in detail the uncharged case $q=0$ (the generalization to $q\neq 0$ is striaghtfoward). The analysis is very similar to the one presented in the previous section so we will be brief, emphasizing mainly the differences. 

As before, we begin by reminding the reader the results for the free energy and entropy without an insertion of the fermion parity operator. In the fixed temperature ensemble we can sum over all states by taking $\Omega=0$ and therefore the black hole solution has $a=0$. The size of the horizon is determined through the (inverse) temperature by 
\beq
\beta = \frac{4\pi r_+}{1+\frac{3 r_+^2}{\ell^2}},~~~r_+ = \frac{2\pi \ell^2}{3\beta}\left( 1 + \sqrt{1-\frac{3\beta^2}{4\pi^2 \ell^2}}\right).
\eeq
This solution only exists for $\beta \leq 2\pi \ell /\sqrt{3}$. The free energy, computed from the on-shell action $I= \beta E - A/4 - \beta \Omega J$ is given by
\bea
-\beta F&=&\frac{(2\pi \ell^2 + \sqrt{4\pi^2 \ell^4 - 3 \beta^2 \ell^2})(-3\beta^2 + \pi (2\pi \ell^2 + \sqrt{4\pi^2 \ell^4 - 3 \beta^2 \ell^2}))}{27\beta^2}\\
&\sim& \frac{16 \pi^3 \ell^4}{27 \beta^2},~~~~\beta\to 0.
\ea
From the first line we can derive that the black hole dominates over thermal AdS for temperatures higher than $\beta < \pi \ell$. The second line shows the result in the high temperature limit. There is a second solution for $r_+$ which has higher free energy and therefore never dominates the ensemble.

In the fixed energy ensemble, the black hole radius is constrained by $E = \frac{r_+(r_+^2 + \ell^2)}{2\ell^2}$. There are three solutions of this equation, one real and two complex conjugate ones. The two complex solutions have negative real part of the entropy. The real solution has an entropy given by 
\beq
S(E) = \frac{\pi(3^{2/3}\ell^2 - 3^{1/3}\ell^{4/3} (\sqrt{3\ell^2+81 E^2}-9 E)^{2/3})^2}{9 \ell^{4/3}(\sqrt{3\ell^2+81 E^2}-9 E)^{2/3}} \sim 2^{2/3} \ell^{4/3} \pi E^{2/3},~~~~E\to \infty,
\eeq
where we also indicated the behavior at large energies. In this limit the radius of the black hole is $r_+ \sim 2^{1/3} \ell^{2/3} E^{1/3}$.

We are now ready to consider the distribution between fermionic and bosonic black hole microstates. We begin by computing the free energy $F_\spin$  in the presence of a fermion parity operator insertion $(-1)^{\sf F}$. Again, we implement this by imposing $\beta \Omega = 2\pi \i$. The first observation we can make is that nothing interesting happens that requires regulating the limit $\beta \Omega \to 2\pi \i$, as opposed to the flat space analysis. We find two solutions for $a=\i {\sf a}$ given by 
\beq\label{eq:AdS4aa}
{\sf a} = r_+ ,~~~~{\sf a} = \frac{r_+(\ell^2 + 3 r_+^2)}{\ell^2 -r_+^2}.
\eeq
The first solution has $\beta=0$ and $E=0$ so it is the same as ${\sf a}=r_+$ in flat space. The second solution has temperature and energy
\beq\label{eq:AdS4bE}
\beta = \frac{16 \pi \ell^2 r_+^3}{3r_+^4 - 2 \ell^2 r_+^2 -\ell^4},~~~E= - \frac{4 r_+^3 \ell^2 (\ell^2 - r_+^2)^2}{(9 r_+^4 - 2 \ell^2 r_+^2 +\ell^4)^2}
\eeq
To compute $F_\spin$ we focus on the first relation and solve for $r_+(\beta)$. For general $\beta$ and $\ell$, we find four solutions for $r_+$. Two are real and two are complex conjugated, and they are all finite. In the flat space limit, one of the solution for the black hole radius becomes the one found in the previous section with $r_+ \to \infty$. We see therefore that the background cosmological constant regulates the large $r_+$ limit of the flat space black hole, without the need to go away from $\beta \Omega = 2\pi \i$. We find that all these four solutions satisfy, for any $\beta$ and $\ell$,
\beq
{\rm Re}\left[ - \beta F_\spin \right] < 0,
\eeq
and therefore thermal AdS always dominates the ensemble computing $F_\spin$. The possible presence of other saddles which are neither thermal AdS nor black holes and support periodic fermion boundary conditions is not ruled out. Assuming this is not the case and the free energy is indeed of order one, controlled by thermal AdS, it suggests a large cancellation in $F_\spin$ compared with $F$ since there are actually a large number of black hole states in a given energy window.

\smallskip 

Next we consider the entropy $S_\spin$, counting black hole states in the presence of a $(-1)^{\sf F}$ insertion. Using equation \eqref{eq:AdS4bE} we solve for $r_+(E)$. It is clear by looking at the equation for $E$ that it can be rewritten as the roots of an eight order polynomial. There are therefore eight solutions (that cannot be written analytically). We find for positive values of the energy that the solutions always come in four pairs of complex conjugated values of $r_+$. Out of these four pairs, two pairs have negative real part of the action while two pairs have positive real part of the action. We denote the action of these solutions by $I_1$ and $I_2$, such that ${\rm Re}(I_{1,2})<0$ and ${\rm Im}(I_{1,2})>0$, and their complex conjugate. Out of the two pairs with negative real part of the action, one pair has the smallest real part which we choose to be $I_2$.

In general, in the presence of two competing saddles with the same real part we should add their contributions in the total partition function. The one-loop determinants arising from each saddle are important in writing the total contribution. In our case the two saddles are related by $\i \to -\i$ and therefore the one-loop determinants are related by complex conjugation. The total answer is
\beq
Z \approx Z_{\rm 1-loop} \,e^{-I} + \overline{Z_{\rm 1-loop}} \,e^{-\bar{I}} \approx |Z_{\rm 1-loop} | e^{- {\rm Re}(I)} \,2\cos\left( {\rm Im}(I)+{\rm Arg}\,Z_{\rm 1-loop}\right).
\eeq 
This applies whether $Z$ computes the partition function in the fixed energy or temperature ensemble. As a function of either, the real part of the action and the absolute value of the one-loop determinant determine the envelope of the partition function while the imaginary part of the action leads to rapid oscillations around the envelope. The overall phase of these oscillations requires knowledge of the one-loop determinants which we do not evaluate here, but the frequency to leading order is determined by ${\rm Im}(I)$. A similar feature takes place in the context of the WKB approximation in quantum mechanics. 
This conclusion applies to both $I_1$ and $I_2$.

The pair of solutions with minimal ${\rm Re}(I)$, with an action $I_2$, would naively be the one that dominates the ensemble. Nevertheless, we find it to have strange properties. For example, it satisfies ${\rm Re}(r_+)<0$ \footnote{One can check that they do not satisfy the Kontsevich-Segal-Witten criterion \cite{Kontsevich:2021dmb,Witten:2021nzp}, see Appendix \ref{app:KSW} for more discussion.} and moreover if we continue this solution smoothly from $\beta\Omega= 2\pi \i$ down to $\beta \Omega = 0$ it does not become the familiar black hole that computes the partition function and instead remains complex. On the other hand the other solution with action $I_1$ is less dominant but has a more reasonable behavior. For example, ${\rm Re}(r_+)>0$ and becomes the black hole computing the partition function when continued to $\beta \Omega=0$. Moreover, only this solution becomes the one studied in Section \ref{sec:4dflat} in the flat space limit. For these reasons we conjecture that the integration contour in the gravitational path integral is such that the solutions $I_2$ do not contribute and instead the pair of complex conjugate solutions with action $I_1$ and ${\rm Re}(r_+)>0$ dominate the ensemble.
\begin{figure}
\begin{center}
\hspace{-0.3cm}\includegraphics[scale=0.55]{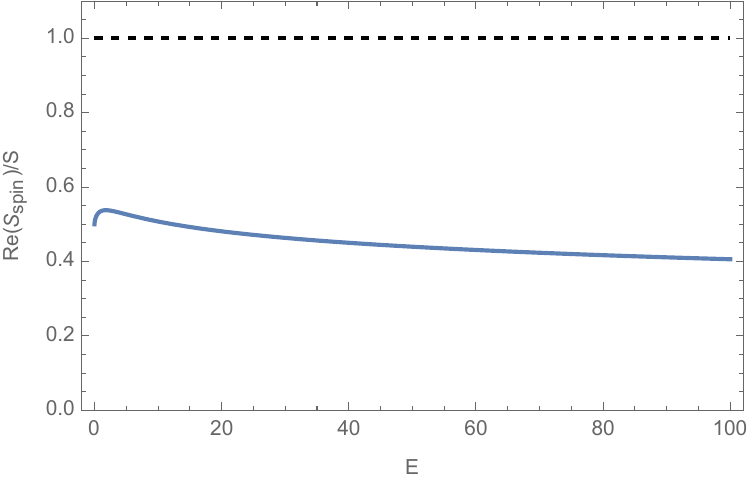}\hspace{1cm}
\includegraphics[scale=0.55]{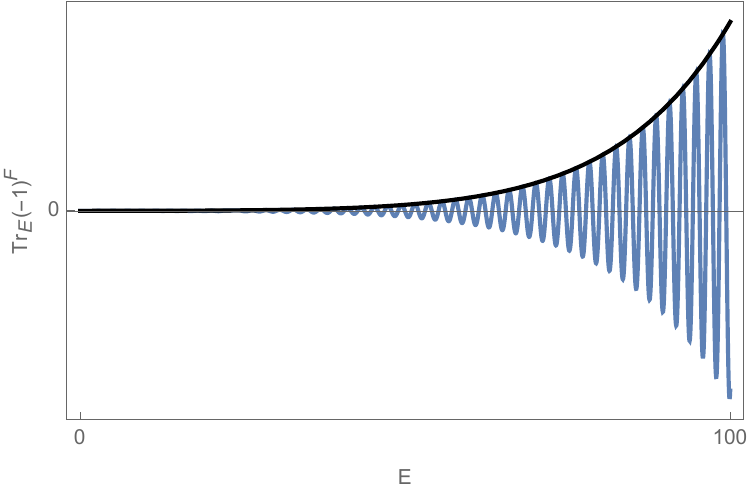}
\end{center}
\vspace{-5mm}
\caption{\footnotesize \textit{Left:} Plot of the envelope of $S_\spin/S$ for a black hole in AdS${}_4$ with $\ell=10$ as a function of energy $E$, in blue. It starts at $1/2$ (flat space result), grows slightly at small energy and then decay to zero at large energy. \textit{Right:} Sketch - not to scale - of the rapid fluctuations as a function of energy in ${\rm Tr}_E \, (-1)^{\sf F}$ in blue and the envelope (black) growing exponentially with $\sqrt{E}$. Even though we can trust the frequency of oscillations, the overall phase is sensitive to quantum corrections.}
\label{fig:4dads}
\end{figure}

To get some insight on these results we can find the black hole solution analytically in the large energy limit. Looking at equation \eqref{eq:AdS4bE} the solution for large $E$ is such that the denominator vanishes $9 r_\star^4 - 2\ell^2 r_\star^2 + \ell^4=0$. (As opposed to the partition function without a $(-1)^{\sf F}$ insertion, large $r_+$ leads to small energies.) This equation has four solutions $r_\star = \pm \frac{\ell}{3} \sqrt{ 1\pm \i 2^{3/2} }$ for each choice of signs. Then we expand $r_+ = r_\star + \delta r$ and solve for $\delta r$ as a function of $E$. For large $E$, $\delta r$ will be small, so we can solve for $\delta r$ perturbatively. The solution we propose to keep is the one that arises from a small fluctuation around $r_\star = \frac{\ell}{3} \sqrt{ 1\pm \i 2^{3/2} } \approx 0.47\ell \pm \i 0.33\ell$. The other solutions have negative real part for $r_\star$. In this limit the action is given by 
\beq
I_1 =  c  \sqrt{\ell^3 E} + 2\pi \i \ell E + \ldots,~~~~c\approx 3.81 ,
\eeq
where the dots denote terms subleading in the large energy limit. The other saddle in the pair has action $\bar{I}_1=c \sqrt{\ell^3 E} - 2\pi \i \ell E$. Since these two solutions have the same real part of the action they both contribute and should be added up in the path integral. The result for the entropy in the presence of $(-1)^{\sf F}$ is therefore 
\beq
S_\spin(E) \approx c \sqrt{ \ell^3 E} + \log  \cos(2\pi \ell E)
\eeq
in the large energy limit. As explained above, knowing the actual phase of the oscillatory term requires including quantum effects, a problem we leave for future work. This result explains the reason why the black hole states do not contribute to the canonical ensemble with the insertion of $(-1)^{\sf F}$. We find within each energy window the difference between bosonic and fermionic states is large in magnitude, of order $e^{c \sqrt{\ell^3 E}}$.\footnote{This expression looks like the Cardy formula in 2d and would be interesting if there was a two dimensional interpretation of this result from the CFT side. It might very well be a coincidence.} Nevertheless whether most states are bosonic or fermionic depends also on the energy due to the oscillatory factor coming from the imaginary part of the action, making the sum over energies (involved in the fixed temperature ensemble) smaller than each individual term. 

\bigskip 

The contribution to the difference between fermionic and bosonic states, coming from the black hole studied here, is qualitatively consistent with \cite{Cherman:2018mya}. The authors of \cite{Cherman:2018mya} propose that without supersymmetry, large $N$ adjoint QCD in four dimensions has a large cancellation between bosonic and fermionic states. The partition function with an insertion of $(-1)^{\sf F}$ is found to grow with energy, with a rate slower than the partition function. This is similar to the prediction we find here for three dimensional conformal field theories.

\subsubsection{AdS${}_5$}

In this section we make some comments regarding the generalization to five dimensional AdS, relevant to four dimensional conformal field theories. We consider a low energy theory with an action given by Einstein-Hilbert with cosmological constant $\Lambda = -6/\ell^2$, with $\ell$ being the five-dimensional AdS radius. For simplicity we consider the case without charge. The black hole metric with mass and angular momentum is given by
\bea
\d s^2 &=& - \frac{\Delta_\theta (1+\frac{r^2}{\ell^2})}{\Xi_a \Xi_b} \d t^2 + \frac{2 m}{\rho^2}\left(\frac{\Delta_\theta \d t}{\Xi_a \Xi_b} - \omega\right)^2 + \frac{\rho^2 \d r^2}{\Delta_r} + \frac{\rho^2 \d \theta^2}{\Delta_\theta} \nonumber\\
&&+ \frac{r^2+a^2}{\Xi_a} \sin^2\theta \d \phi^2 + \frac{r^2+b^2}{\Xi_b} \cos^2\theta \d \psi^2.
\ea
where we define $\rho^2 = r^2 + a^2 \cos^2 \theta +b^2 \sin^2 \theta$ and the functions 
\bea
&& \Xi_a = 1-\frac{a^2}{\ell^2},~~~~\Xi_b=1-\frac{b^2}{\ell^2},~~~~~\omega = a \sin^2\theta \frac{\d \phi}{\Xi_a} + b \cos^2 \theta \frac{\d \psi}{\Xi_b},\nonumber\\
&& \Delta_r = \frac{(r^2 + a^2)(r^2 +b^2)(1+\frac{r^2}{\ell^2})}{r^2}-2m,~~\Delta_\theta = 1-\frac{a^2}{\ell^2} \cos^2 \theta -\frac{b^2}{\ell^2} \sin^2 \theta.
\ea
The boundary has the topology of $S^1 \times S^3$ with $S^1$ being thermal time. The spatial three-sphere is parametrized by $\phi$ and $\psi$ which are $2\pi$ periodic and $\theta \in [0,\pi/2]$. When considering a solution with a Lorenzian interpretation one has to restrict $a^2<\ell^2$ and $b^2 < \ell^2$. In the Euclidean path integral involved in computing the free energy or entropy, we allow these parameters to be complex. The two angular momenta $J_1$ and $J_2$ generate rotations around $\phi$ and $\psi$ respectively. The energy and angular momenta for this black hole are given by 
\beq
E = \frac{\pi m (2 \Xi_a + 2 \Xi_b - \Xi_a \Xi_b)}{4 \Xi_a^2 \Xi_b^2},~~J_1 = \frac{\pi 2 a m}{4 \Xi_a^2 \Xi_b},~~J_2 = \frac{\pi 2 b m}{4 \Xi_a \Xi_b^2}.
\eeq
The thermodynamic potentials, the temperature and the two angular velocities conjugated to the two angular momenta $J_1$ and $J_2$, are derived from imposing smoothness at the Euclidean horizon and are given by
\beq
\beta = \frac{2\pi r_+ (r_+^2 +a^2)(r_+^2 +b^2)}{r_+^4[1+\ell^{-2}(2r_+^2+a^2+b^2)]},~~\Omega_1 = \frac{a (1+\frac{r_+^2}{\ell^2})}{(r_+^2 + a^2)},~~~\Omega_2 = \frac{b (1+\frac{r_+^2}{\ell^2})}{(r_+^2 + b^2)}.
\eeq
The area of the event horizon that appears in the Bekenstein-Hawking entropy is given by $A = \frac{\pi^2 (r_+^2 + a^2 )(r_+^2 + b^2)}{2 r_+ \Xi_a \Xi_b}$. The on-shell action in the fixed temperature and angular velocity ensemble is $I= \beta E - A/4 - \beta \Omega_1 J_1 - \beta \Omega_2 J_2$. In the fixed energy and angular velocity ensemble, the on-shell action is instead $I=  - A/4 - \beta \Omega_1 J_1 - \beta \Omega_2 J_2$.

In the fixed temperature ensemble we can compute the free energy with no insertion of $(-1)^{\sf F}$. The saddle point in this case contributing to the partition function is a black hole with $a=b=0$ and $r_+$ is computed as a function of inverse temperature $\beta$. The solution is only real for $\beta < \pi \ell /2$. Moreover the black hole solution dominates the ensemble only for even higher temperatures $\beta < 2 \pi \ell /3$. In the range $2\pi \ell/3 < \beta $ thermal AdS dominate the ensemble and therefore $\beta_{\rm HP} = 2 \pi \ell/3$ is the Hawking-Page transition corresponding to the confinement-deconfinement transition of the dual gauge theory \cite{Witten:1998zw}. At very high temperatures, the size of the black hole is $r_+\approx \ell^2\pi/\beta$ and the free energy is large and negative
\beq
- \beta F \approx \frac{\ell^6 \pi^5}{8 \beta^3}
\eeq
and in a fixed energy ensemble the entropy grows as $S(E) \sim E^{3/4}$. 

\bigskip 

We now move on to the computation of $F_\spin$ and $S_\spin$. In this case (and really for black holes in any number of dimensions bigger than four) we have more freedom to decide how to implement the $(-1)^{\sf F}$ since there are more than one angular momenta. In our definition $J_1$ and $J_2$ are two generators that each rotate along an $\mathbb{R}^2 \subset \mathbb{R}^4$. From the point of view of the field theory living in the four dimensional boundary of AdS${}_5$,  Lorentz invariance on $\mathbb{R}^4$ implies that $J_1$ and $J_2$ are either both half-integer (corresponding to fermionic states) or both integer (corresponding to bosonic states). In other words $2J_1 = 2J_2~{\rm mod}~2\mathbb{Z}$. Therefore we could implement the fermion parity operator either by $(-1)^{\sf F} = e^{2\pi \i J_1}$ or $(-1)^{\sf F} = e^{2\pi \i J_2}$. For concreteness, we choose to use $J_1$ and therefore we can set $\Omega_2 = J_2 = 0$ on our black hole.

There are other notions of fermion parity operator in five dimensions. The group of rotations of $S^3$ can be written as ${\rm SO}(4) \approx {\rm SU}(2)_L \times {\rm SU}(2)_R$ and the two ${\rm SU}(2)$ factors are generated by $J_L = \frac{J_1+J_2}{2} $ and $J_R=\frac{J_1-J_2}{2}$. Therefore one could also define other parity operators such as $(-1)^{{\sf F}_{\sf L}}= e^{2\pi \i J_L}$ and $(-1)^{{\sf F}_{\sf R}}= e^{2\pi \i J_R}$. The free energy and entropy in the presence of such insertions can be computed in a similar way, but they would not be probing the quantum statistics of the black hole microstate which requires the  $(-1)^{\sf F}$ insertion of the previous paragraph. 

\smallskip 

After these clarifications we implement the insertion of $(-1)^{\sf F}$ by setting $\beta \Omega_ 1 = 2\pi \i $ and $\beta \Omega_2 =0$ for concreteness. The second condition implies $b=0$ and the solution for $a=\i {\sf a}$ is 
\beq\label{solnads5}
\frac{\beta \Omega_1}{2\pi \i} = \frac{{\sf a}(r_+^2+\ell^2)}{2r_+^3-{\sf a}^2 r_+ +\ell^2 r_+}=1 ,~~\Rightarrow~~{\sf a} = -\frac{\ell^2}{r_+} - 2 r_+.
\eeq
(There is also another solution with ${\sf a} =r_+$ and with $\beta=E=0$ which we ignore.) With (\ref{solnads5}), the temperature and energy are given by
\beq
\beta = \frac{2\pi \ell^2 (3 r_+^2+\ell^2)}{2 r_+^3 +\ell^2 r_+},~~~~E = - \frac{\pi(3 r_+^2+\ell^2)(4 r_+^4 + 7\ell^2r_+^2 +\ell^4)}{8(4r_+^2+\ell^2)^2}.
\eeq
In the fixed temperature ensemble we can solve $r_+(\beta)$ and compute the free energy $-\beta F_\spin = -\beta E + A/4 + \beta \Omega_1 J_1$. In a fixed energy ensemble we solve instead $r_+(E)$ and $S_\spin = A/4 + \beta \Omega_1 J_1$.

\bigskip 

In the fixed temperature ensemble, we find numerically that all solutions for $r_+$ lead to a free energy ${\rm Re}[-\beta F_\spin ] <0 $ and therefore thermal AdS${}_5$ dominates the ensemble for $F_\spin$ at all temperatures. There is no Hawking-Page transition. This is similar to the results for AdS${}_4$, and it is also qualitatively consistent with \cite{Cherman:2018mya}. From the boundary side the Hawking-Page transtition present in the free energy is interpreted as a confinement-deconfinement phase transtition. On the other hand the free energy in the presence of $(-1)^{\sf F}$ has no phase transition and is always in the confined phase regardless of temperature.  

\bigskip 

In the fixed energy ensemble we find numerically that ${\rm Re}[S_\spin]=0$ for all the solutions for $r_+$. Therefore, as opposed to the result in four dimensions, $S_\spin$ is controlled by the difference between bosonic and fermionic low energy excitations around thermal AdS, and not from a black hole solution. In this case we expect any type of bulk spin defect that supports periodic fermions to dominate as long as it has large enough entropy. We will come back to this later in the discussion section.

\subsubsection{AdS${}_3$} 
We now consider the case of black holes in asymptotically AdS${}_3$ spaces. We focus on theories with a gravity sector described by the Einstein-Hilbert action with a negative cosmological constant, with the AdS${}_3$ length given by $\ell$. The black hole solution in this theory is given by the BTZ metric
\beq
\d s^2 = - f \d t^2 + \ell^2 \frac{\d r^2}{f} + r^2 (\d \phi - \frac{r_- r_+}{r^2}\d t)^2,~~~f =\frac{(r^2 -r_+^2)(r^2 - r_-^2)}{r^2}.
\eeq
This black hole carries energy and angular momentum, given by
\beq
E = \frac{\ell}{8} + \frac{r_+^2 + r_-^2}{8 \ell} ,~~~J = \frac{r_+ r_-}{4 \ell}.
\eeq
To be consistent with the other sections, we define the energy such that vacuum AdS has $E=0$, which is an unconventional choice in the context of 2d CFTs. The corresponding inverse temperature and angular velocity are given by
\beq
\beta  = \frac{2 \pi \ell r_+}{r_+^2 - r_-^2},~~~~~\Omega = \frac{r_-}{r_+}.
\eeq
The area of the horizon is $A= 2 \pi r_+$. It is convenient to work with the left and right-moving temperatures which are $\beta_L = \beta + \beta \Omega$ and $\beta_L = \beta - \beta \Omega$. The on-shell action of the BTZ black hole is given, in an ensemble of fixed left- and right-moving temperatures, by $I= \beta E - A/4 - \beta \Omega J$, while the action of vacuum AdS is just zero. The contribution of the BTZ black hole in a fixed temperature and angular velocity ensemble to the free energy is 
\beq
-\beta F= - \frac{\beta_L c}{24} + \frac{c}{24} \frac{4\pi^2}{\beta_L} - \frac{\beta_R c}{24} +\frac{c}{24} \frac{4\pi^2}{\beta_R},
\eeq
where $c=3 \ell/ 2$. We are working with units where $G_N=1$ and therefore the semiclassical limit corresponds to large AdS radius in Planck units, or $c \gg 1$.

We shall begin by considering the free energy without and with the $(-1)^{\sf F}$ insertion. We always work in the NS sector, meaning that fermions are antiperiodic along the spatial circle.\footnote{Notice that otherwise the trick explained in figure \ref{fig:ident} to produce black hole solutions with periodic fermions in the thermal circle would not work. See the end of the section for comments on the case of the elliptic genus.} If $\beta \Omega = 0$ or $2\pi \i$ we obtain respectively
\beq
-\beta F = - \frac{\beta c}{12}+ \frac{c}{12} \frac{4\pi^2}{\beta} ,~~~~ - \beta F_\spin = - \frac{\beta c}{12}+\frac{c}{12} \frac{4\pi^2 \beta}{\beta^2 + 4\pi^2}.
\eeq
From the expression for $F$ we see that the black hole only contributes for $\beta < 2\pi$, when compared with thermal ${\rm AdS}_3$. Instead, $-\beta F_\spin < 0 $ for any value of the temperature. This means that thermal AdS with periodic fermions always dominates over the rotating black hole introduced here, in the canonical ensemble. This situation is similar to AdS in other dimensions.

When considering the gravitational path integral in asymptotically AdS${}_3$ spaces one needs to sum over ${\rm SL}(2,\mathbb{Z})$ images where different cycles are contractible at the horizon, but satisfies the same boundary conditions \cite{Maldacena:1998bw}. Since we are working with theories with fermions we need to keep track of the spin structure and therefore the sum is only over the subgroup of ${\rm SL}(2,\mathbb{Z})$ that does not modify the spin structure.\footnote{In the NS sector this involves a sum over the subgroup $\Gamma_\theta/\mathbb{Z} \subset {\rm SL}(2,\mathbb{Z})$. The group $\Gamma_\theta$ is of the form $\Big(\begin{matrix}
	a & b \\ c &d
\end{matrix}\Big)$ with both $a+b$ and $c+d$ odd \cite{Maloney:2007ud}, and we mod out by the group generated by $\Big(\begin{matrix}
	1 & 2 \\ 0 & 1
\end{matrix}\Big)$.}  For the thermal trace with $\Omega=0$ it is known that the BTZ black hole  dominate the ensemble (or, at low enough temperatures, thermal AdS). For $\Omega= 2\pi \i$ we have checked that there is no other ${\rm SL}(2,\mathbb{Z})$ image (other than thermal AdS) that  respects the spin structure and produces a more dominant contribution than the rotating black hole. This is not obvious since, for example, at $\Omega = \pi \i$ there is an ${\rm SL}(2,\mathbb{Z})$ black hole whose contribution is bigger than either thermal AdS or the analog of our rotating black hole, and actually dominates the ensemble, see discussion in \cite{MYITP:2023sep}.\footnote{The authors of \cite{MYITP:2023sep} describe an orbifold of the BTZ black hole that makes the dominant contribution. We checked the action of this orbifold is precisely one of the ${\rm SL}(2,\mathbb{Z})$ images of the rotating black hole with $\Omega = \i \pi$.} This does not happen for our observable involving $\Omega = 2\pi \i$.

\smallskip

We now analyze the difference between bosonic and fermionic black holes in the fixed energy ensemble. The result for the total black hole entropy is the well-known Cardy formula $S(E) = 2\pi \sqrt{\frac{c}{3} (E-\frac{c}{12})}$ for $E\geq \frac{c}{12}$. (Only for $E\geq c/6$ the BTZ black hole also dominates the fixed temperature ensemble \cite{Hartman:2014oaa}.). What is the gravity prediction for $S_\spin$? The procedure is very similar to the one implemented already so we will be brief. Setting $\beta \Omega= 2\pi \i$, we find eight solutions for $r_+$ given $E$: four pairs of complex conjugated ones. Two pairs have ${\rm Re}[ S_\spin]\leq 0$, and one pair has ${\rm Re} [S_\spin] >0$. The pair with positive entropy gives the leading contribution but they come with an imaginary part of opposite sign, leading to a similar oscillatory term as in AdS${}_4$. The result is 
\beq
{\rm Re} \,[S_\spin (E)] = \sqrt{2} \pi \sqrt{\frac{c}{3} E} = \frac{1}{\sqrt{2}} S(E),~~~~E\gg c.
\eeq
and ${\rm Im}\,[S_\spin] = \pm 2 \pi   E$. Therefore even though we expect the entropy in the presence of $(-1)^{\sf F}$ to be large, the cancellation found in the fixed temperature ensemble is due to the rapid oscillations as a function of energy. This situation is completely analogous to ${\rm AdS}_4$. 

\bigskip 

We would like to emphasize that $(-1)^{\sf F}$ is defined as $e^{2\pi \i J}$ and therefore counts the difference between black hole microstates with integer or half-integer scaling dimensions. This is different than the elliptic genus usually computed in two dimensional supersymmetric theories. The elliptic genus counts the difference between states with even vs. odd charges with respect to a ${\rm U}(1)$ gauge field (which is normally the Cartan of a bigger non-Abelian group such as ${\rm SU}(2)$). When AdS${}_3$ arises form string theory, this charge comes from angular momentum on a compact direction and therefore can also be interpreted as a different version of a fermion parity operator similar to the discussion about ${\rm AdS}_5$. Since in this case the smoothness of the choice of spin structure is thanks to the presence of a background gauge field (and not 3d rotation) the procedure of implementing the fermion parity operator through a complex chemical potential works also in the R sector.

\section{Contribution from wormholes}\label{sec:wormhole}

One might wonder whether wormholes contribute to the quantity $\rho_{\textrm{spin}} (E) \equiv \textrm{Tr}_{E} (-1)^{\sf F}$. To be more precise, what we mean is that the semiclassical geometry we discussed in the previous sections only give a coarse-grained estimation to the quantity $\rho_{\textrm{spin}} (E)$, which varies smoothly with the energy $E$, in spite of the oscillation at the scale of $1/\ell$. However, in a microscopic theory, one expects that on top of the smoothly varying average, $\rho_{\textrm{spin}} (E)$ also contains a fluctuating piece that depends sensitively to the precise energy window one chooses. We can quantify the typical size of the fluctuating piece by studying the connected correlator
\begin{equation}\label{corr}
	\langle \rho_{\textrm{spin}} (E) \rho_{\textrm{spin}} (E') \rangle_\textrm{c} \equiv \langle\rho_{\textrm{spin}} (E) \rho_{\textrm{spin}} (E') \rangle -\langle \rho_{\textrm{spin}} (E) \rangle\langle \rho_{\textrm{spin}} (E')  \rangle .
\end{equation}
Here the bracket $\langle \cdot\rangle$, in a fixed microscopic theory, can be simply viewed as averaging over a small range of energy $E$ while keeping $E-E'$ fixed. In a theory with only bosons, the correlator (\ref{corr}) is nothing but the correlator of the full density of state $\rho(E)$. Here we are interested in theories that have fermions, where the bosonic and fermionic states contributes to $\rho_{\textrm{spin}} (E)$ with different signs. 

Following the insight in \cite{Penington:2019kki,Saad:2018bqo,Stanford:2020wkf}, the connected contribution to the average (\ref{corr}) can be quantified by looking at the contribution from wormholes in the gravitational path integral. 
In supersymmetric theories, however, it has been argued in \cite{Iliesiu:2021are} that wormholes do not contribute to the square of the index. However, there is no argument in non-supersymmetric theories, which we are focusing on here.

In this section, we discuss a universal wormhole contribution to a closely related quantity
\begin{equation}\label{sssf}
	\langle |Y_{E,\textrm{spin}} (T)|^2 \rangle_\textrm{c} \equiv \langle |\textrm{Tr}_{E} \, (-1)^{\sf F} e^{-\i HT}|^2 \rangle_{\textrm{c}}
\end{equation}
where we introduced a factor of unitary time evolution $e^{-\i HT}$ into the trace other than the original $(-1)^{\sf F}$ factor. Evidently, this is the analogue of the spectral form factor \cite{Cotler:2016fpe,Saad:2018bqo} when we don not have the $(-1)^{\sf F}$ insertion and we will therefore call it the spin spectral form factor. (A similar quantity was introduced for supersymmetric theories in \cite{Choi:2022asl}.)

\begin{figure}
\begin{center}
\includegraphics[scale=0.17]{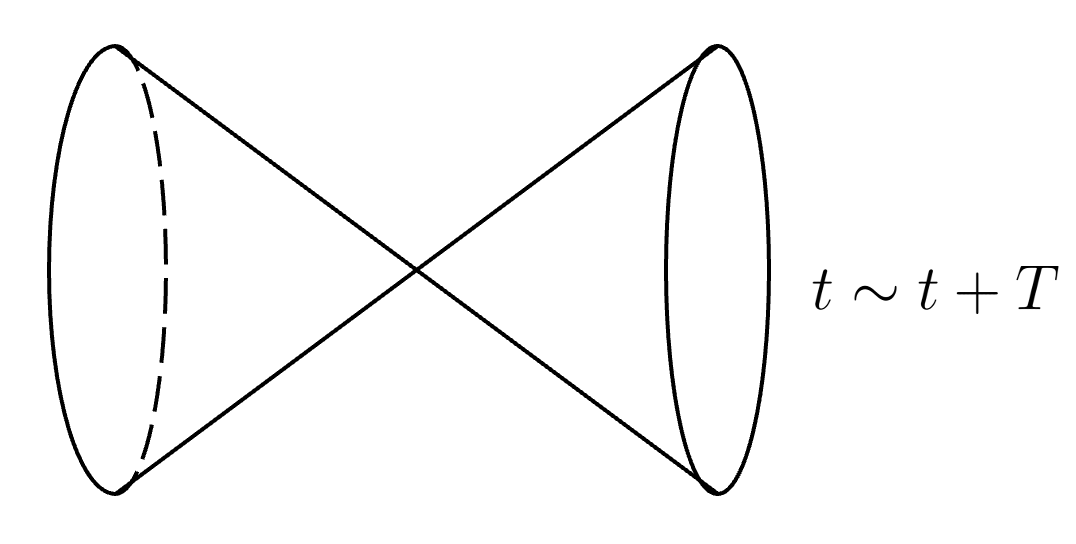}
\end{center}
\vspace{-5mm}
\caption{\footnotesize An illustration of the double cone wormhole geometry, where we are only drawing the radial and time directions.}
\label{fig:doublecone}
\end{figure}

The wormhole contribution to the ordinary spectral form factor
\begin{equation}\label{ossf}
	\langle |Y_{E} (T)|^2 \rangle_\textrm{c} \equiv \langle |\textrm{Tr}_{E} \, e^{-\i HT}|^2 \rangle_{\textrm{c}}
\end{equation}
was discussed in \cite{Saad:2018bqo}. In particular, in the microcanonical ensemble where one focuses on states around energy $E$, the gravity solution is the double cone wormhole which exists universally in any theories containing black holes. To construct it, one starts with a two-sided black hole geometry with energy $E$ and perform a quotient in time by period $T$, resulting in a geometry of the form in figure \ref{fig:doublecone}. The resulting geometry has a zero mode corresponding to the relative time shift between the two sides, which leads to a linear $T$ dependence in the spectral form factor.  The geometry is naively singular since the time circle shrinks to zero size at the bifurcation surface, while \cite{Saad:2018bqo} proposed a simple prescription to regulate the geometry by deforming the radial coordinate slightly into the complex plane. The time circle is non-shrinking everywhere in the geometry. Another feature of the double cone is that its classical action is zero due to the cancellation between the left and right sides. Since the action vanishes, black holes carrying different angular momentum and possible gauge charges contribute equally (the coefficient multiplying $T$ in the one-loop determinant is charge independent as well), and one should sum over them. In the case without any symmetry and the statistics of eigenvalues follow Gaussian unitary ensemble (GUE) universality, we have\footnote{Any unitary Lorentz invariant theory comes with an additional symmetry that modifies the statement here. See footnote \ref{footnotecpt} for more discussion. }
\begin{equation}\label{eq:SSFTE}
    \langle |Y_{E} (T)|^2 \rangle_\textrm{c} \approx  \int \d E\, \frac{T}{2\pi},
\end{equation}
where the integral over $E$ runs over the energy window involved in defining the spectral form factor.

To understand the computation of (\ref{sssf}), we first discuss an analogous situation where we have a $\U(1)$ gauge symmetry which is also interesting in its own right.

\subsection{Analogue in the case with a $\U(1)$ gauge symmetry}\label{sec:U1}

In a gravity theory with a $\U(1)$ gauge symmetry, we can consider an analogue of the quantity (\ref{sssf}) as
\begin{equation}\label{Ycharge}
	\langle |Y_{E,\mu} (T)|^2 \rangle_\textrm{c} \equiv \langle |\textrm{Tr}_{E} \, e^{-\i (H - \mu Q )T}|^2 \rangle_{\textrm{c}}, \quad \mu\in \mathbb{R}
\end{equation}
namely we weight the states carrying charge $Q$ by a pure phase factor $e^{\i \mu QT}$. In gravity, the way to implement this phase factor is to impose a non-zero boundary condition for the gauge field at infinity, such that the holonomy of the gauge field satisfies
\begin{equation}\label{bdyholo}
	e^{\i \int \d t\, A_t}|_{\textrm{bdry}} = e^{\i\mu T}.
\end{equation} 
In the gauge that $\partial_t A_t = 0$, (\ref{bdyholo}) determines the boundary value of the gauge field up to constant shifts
\begin{equation}
	A_t|_{\textrm{bdry}} = \mu + \frac{2\pi n}{T} , \quad n \in \mathbb{Z}. 
\end{equation}
Let us focus on the $n=0$ case in the following. The sum over the shifts by $n$ is important for ensuring the quantization condition for the $\U(1)$ charge, but it is not the main focus here (see Appendix. \ref{app:JT} for an explicit example of how it works). 
We have $A_t|_{\textrm{bdry}} = \mu$ and we look for the analogous double cone geometry satisfying the boundary condition.  The immediate guess will be to take the Lorentzian two-sided charged black hole solution, with energy given by $E$ and charge $q$ set by chemical potential $\mu$ through the usual relation in black hole thermodynamics. For example, in four flat spacetime dimensions, we would be tempting to take
\begin{equation}\label{qmu}
	q = 4\pi \mu r_+ 
\end{equation} 
where $r_+$ is the radius of the outer horizon. 
However, even though what is described above will be a valid double cone geometry that contributes to (\ref{Ycharge}), it is only one out of an entire family of solutions. In fact, for the double cone geometry, even after we fixed the boundary value for the gauge field, the charge $q$ can still freely vary instead of being fixed by (\ref{qmu}) or analogous formulae in other dimensions. To understand this point, we can first ask how (\ref{qmu}) could be derived if one were to consider the standard black hole thermodynamics. There the geometry in question is a Euclidean black hole, with the classical solution of gauge field being
\begin{equation}
	A_{t_{\rm E}} = - \mu  + \frac{ q}{4\pi r}
\end{equation}
and the condition (\ref{qmu}) comes from requiring that 
$A_{t_{\rm E}}|_{r=r_+} = 0$, namely the gauge field configuration is smooth at the horizon, where the time circle shrinks to a point.
However, for the double cone geometry, the crucial difference is that the time circle never shrinks in the geometry, so we no longer have the constraint $A_{t}|_{r=r_+} = 0$. Saying it in a different way, the coefficient of the $1/r$ piece  in the gauge field - the physical charge $q$ - is not determined by $\mu$ and can freely vary.

Physically, the fact we no longer have a map between $q$ and $\mu$ for the double cone geometry is reasonable based on the expectation that no particular value of charge $q$ would dominate (\ref{Ycharge}). The same phenomena takes place in the original double cone with respect to the mass of the black hole: the time cycle is not contractible when periodically identifying time on the Lorentzian geometry. Therefore the mass is not fixed by the asymptotic boundary length and the double cone contributes equally at all energies. Indeed, the double cone geometry has zero classical action regardless of the value of $q$, so black holes with all possible charges $q$ contribute equally to (\ref{Ycharge}) at the classical level. In other words, we have
\begin{equation}
    \langle |Y_{E,\mu} (T)|^2 \rangle_\textrm{c} \approx \int \d E  \int \d q\, \frac{T}{2\pi}.
\end{equation}

For the specific case of JT gravity coupled to a $\U(1)$ gauge field, the wormhole contribution has been studied in \cite{Kapec:2019ecr}. In Appendix \ref{app:JT} we review the calculation in JT gravity.  Above we are simply restating the main features they found in a general setting. See also \cite{Cotler:2022rud} for discussion on closely related wormhole solutions with a $\U(1)$ gauge field.

\subsection{Wormhole contribution to the spin spectral form factor}\label{sec:SSSF}

The lesson in the $\U(1)$ symmetry case can be generalized straightforwardly to the spin spectral form factor (\ref{sssf}). To implement the $(-1)^{\sf F}$, similar to what we discussed in Section \ref{sec:bh}, we turn on an angular potential $\Omega = 2\pi/T$ at infinity. 
 In the discussion of Section \ref{sec:bh} where we were studying black hole solutions with a single boundary, it was important that the black hole has particular values of angular momentum $J$ determined by $\Omega$ such that the spin structure is smooth at the horizon. However, in the case of a wormhole geometry, since the time circle is never shrinking, we do not have any constraints on the angular momentum $J$ of the black hole. This is the analogue of the phenomenon that the charge is not fixed by the chemical potential in the previous section. 

\smallskip 

As a conclusion, the wormhole contributions to the spin spectral form factor are simple. One simply consider all the possible black hole geometries with energy $E$, including those with different angular momentum $J$. The only difference from the ordinary double cone is that we have a periodic boundary condition in time for fermionic fields. This change in boundary condition is invisible at the classical level, but will affect the one-loop fluctuations around the geometry. However, given that the one-loop determinant of the matter fields around the double cone geometries goes to one at sufficiently late time \cite{Saad:2018bqo,CIM},\footnote{On the flip side, the matter fluctuations are important at early time, so what we said in this section does not generalized easily to $\langle | \textrm{Tr}_E (-1)^{\sf F} |^2\rangle_{\rm c}$.} we expect that at  large enough $T$ we have
\begin{equation}\label{eq:SSSFeqSSF}
	 \langle |\textrm{Tr}_{E} \,(-1)^{\sf F} e^{-\i HT}|^2 \rangle_{\textrm{c}} \approx  \langle |\textrm{Tr}_{E} \, e^{-\i HT}|^2 \rangle_{\textrm{c}}. 
\end{equation}
This is the main result of this section. 

\subsection{Interpretation in terms of random matrix universality}
In this section we give an interpretation of the fact that at late times the spectral form factor is equal with or without the $(-1)^{
\sf F}$ insertion, see equation~\eqref{eq:SSSFeqSSF}, in terms of late time quantum chaos and random matrix universality.

To simplify the discussion let us assume that the only symmetry of the random matrix ensemble is the presence of a $\mathbb{Z}_2$ symmetry generated by $(-1)^{\sf F}$. This implies that the Hilbert space can be decomposed into a bosonic and fermionic sector $\H = \H_b \oplus \H_f$ such that both $(-1)^{\sf F}$ and the Hamiltonian can be written as $2\times2$ blocks:
\beq\label{minf}
(-1)^{\sf F}=\left(\begin{array} {c|c}  1 & 0\cr \hline 0 & -1\end{array}\right),~~~~~~H=\left(\begin{array} {c|c}  H_b & 0\cr \hline 0 & H_f\end{array}\right).
\eeq
The appropriate random matrix ensemble therefore is to take $H_b$ and $H_f$ to be statistically independent random matrices, assuming the absence of time reversal symmetry. This situation was studied in the context of JT gravity and in particular the connection with wormholes in \cite{Stanford:2019vob,Kapec:2019ecr}. The fact that the two sectors are statistically independent imply in particular that 
\beq
\left\langle {\rm Tr}_{\H_b} e^{-\beta_1 H}  \,\,{\rm Tr}_{\H_f} e^{-\beta_2 H}  \right\rangle_{\rm c} = 0,
\eeq
which we will use below. Finally, we did not specify yet which ensemble $H_b$ and $H_f$ are drawn from. As explained in section 4 of \cite{Yan:2023rjh}, the ${\sf CPT}$ theorem implies that the bosonic sector is in the orthogonal ensemble (GOE) while fermionic sector is in the symplectic ensemble (GSE).\footnote{Following the notation of \cite{Witten:2023snr}, the ${\sf CPT}$ theorem says that the operator ${\sf R} {\sf T}$ is a symmetry of any unitary Lorentz invariant theory. ${\sf R}$ generates a transformation that reverses the sign of one coordinate (any) and ${\sf T}$ is the time reversal operator defined such that it anticommutes with conserved charges. In Euclidean signature ${\sf RT}$ generates a rotation of $180^{o}$ in a plane determined by ${\sf R}$ and the time direction. The fact that for a full rotation $({\sf RT})^2 = (-1)^{\sf F}$ implies the appropriate ensemble for the bosonic and fermionic sectors. \label{footnotecpt}} This will not affect much the results below. 

Let us first evaluate the partition function. Decomposing the Hilbert space into a bosonic and fermionic sector implies that for each realization of the Hamiltonian we can write
\beq
{\rm Tr}_{\H}\, e^{-\beta H} ={\rm Tr}_{\H_b}\, e^{-\beta H} + {\rm Tr}_{\H_f}\, e^{-\beta H}  .
\eeq
The connected contribution to the product of two partition functions is therefore given by 
\beq\label{eq:rmrmr}
\left\langle {\rm Tr}_{\H} e^{-\beta_1 H} \, \, {\rm Tr}_{\H}e^{-\beta_2 H}  \right\rangle_{\rm c}=\left\langle {\rm Tr}_{\H_b}e^{-\beta_1 H}\,\,{\rm Tr}_{\H_b} e^{-\beta_2 H} \right\rangle_{\rm c} + \left\langle {\rm Tr}_{\H_f} e^{-\beta_1 H}\,\,{\rm Tr}_{\H_f} e^{-\beta_2 H} \right\rangle_{\rm c}
\eeq
where we use that the mixed terms vanish due to the statistically independence of the sectors. The right hand side is a universal quantity in the large rank limit of random matrices but we will not need its details below. 

Next we evaluate the partition function with a $(-1)^{\sf F}$. In terms of the decomposition into bosonic and fermionic states the answer for each realization is now
\beq
{\rm Tr}_{\H}\,(-1)^{\sf F} e^{-\beta H} ={\rm Tr}_{\H_b}\, e^{-\beta H} - {\rm Tr}_{\H_f}\, e^{-\beta H}  .
\eeq
When computing the connected contribution to the product of two partition functions, the minus sign that appears above is irrelevant since the sectors are statistically independent
\bea
\left\langle {\rm Tr}_{\H}(-1)^{\sf F} e^{-\beta_1 H} \,\,  {\rm Tr}_{\H}(-1)^{\sf F} e^{-\beta_2 H}  \right\rangle_{\rm c}  &=&  \left\langle {\rm Tr}_{\H_b} e^{-\beta_1 H}\,{\rm Tr}_{\H_b} e^{-\beta_2 H} \right\rangle_{\rm c}\nonumber\\
&&+ \left\langle {\rm Tr}_{\H_f} e^{-\beta_1 H}\,{\rm Tr}_{\H_f} e^{-\beta_2 H} \right\rangle_{\rm c}.
\ea
Therefore regardless of the details of the right hand side, we find that the answer is identical in both cases with or without the $(-1)^{\sf F}$ insertion:
\beq
\left\langle {\rm Tr}_{\H}(-1)^{\sf F} e^{-\beta_1 H} \,\,  {\rm Tr}_{\H} (-1)^{\sf F} e^{-\beta_2 H}  \right\rangle_{\rm c} = \left\langle {\rm Tr}_{\H} e^{-\beta_1 H} \, \, {\rm Tr}_{\H}e^{-\beta_2 H}  \right\rangle_{\rm c}
\eeq
After analytically continuing in $\beta$ and going to the microcanonical ensemble, this results is equivalent to equation \eqref{eq:SSSFeqSSF} derived as a consequence of the evaluation of the double cone contribution to the gravitational path integral. The late time behavior of either term on the right hand side of \eqref{eq:rmrmr} has the same behavior as the double cone. First, it is a quantity that does not grow with the rank of the matrix and correspondingly the double cone on-shell action vanishes. Second, it has a late time ramp with a linear growth in $T$.

\smallskip 

So far we incorporated the $(-1)^{\sf F}$ as a generator of a $\mathbb{Z}_2$ symmetry but did not take into account the fact that it is embedded in a bigger group of rotations in higher dimensions. At late times when the double cone becomes reliable, this approximation for the random matrix ensemble was good enough to reproduce the relation \eqref{eq:SSSFeqSSF}. Nevertheless the model here so far does not reproduce the correct prefactor of the linear-in-$T$ term on either side of equation~\eqref{eq:SSSFeqSSF}. Equation~\eqref{eq:rmrmr} implies the coefficient of the linear-in-$T$ term for a theory with a $\mathbb{Z}_2$ symmetry is twice the result for a theory without such a symmetry. According to the discussion in the previous section, the spectral form factor gets multiplied not by $2$ but by a divergent factor given by the sum over the possible angular momenta of the black hole used to generate the double cone. This factor is divergent but can be regulated by working in an ensemble where the angular momentum is specified in a finite window (which would be the natural way to generalize fixing an energy window that regulates the integral over energies in \eqref{eq:SSFTE}). 

A more accurate ensemble incorporates the rotation group $G$ (for example $G=\SU(2)$ in four dimensions or $G=\SU(2) \times \SU(2)$ in five dimensions). The Hilbert space is decomposed into irreducible representations of $G$. For example the case relevant to four dimensions is $\H = \oplus_{J=0,\frac{1}{2},1,\ldots} \H_J $, where each $\H_J$ represents an irreducible $(2J+1)$-dimensional representation of $G=\SU(2)$.\footnote{Here $J$ labels the eigenvalue of $\vec{J}^2$ which is fixed within each representation. In the previous sections we used $J$ to denote the eigenvalue of $\vec{J} \cdot \vec{n}$ instead, the eigenvalue of the angular momentum itself along a given direction.} Since the Hamiltonian commutes with the generators of $G$, it is block diagonal according to the decomposition of the Hilbert space into sectors transforming in fixed representations, and consists of statistically independent random matrices in each sector. We can embed the $\mathbb{Z}_2$ group generated by $(-1)^{\sf F}$ inside $G$.  Moreover, now the ${\sf CPT}$ theorem implies the existence of an ${\sf RT}$ symmetry that squares to $(-1)^{\sf F}$ and in even dimensions (after possibly combining with a rotation) anticommutes with the generator of rotations $\vec{J}$, so even-spin sectors are GOE and odd-spin are GSE.\footnote{This is true for $\SU(2)$ since all representations are real. If the group $G$ has complex representations then such sectors would remain GUE.}  

An insertion of $(-1)^{\sf F}$ introduces a minus sign in the partition function for all representations of $G$ that are odd under $(-1)^{\sf F}$ (for example half-integer $J$'s for $G=\SU(2)$). Since different representations are statistically independent, this sign disappears when evaluating the spectral form factor for essentially the same reason as the $\mathbb{Z}_2$ case studied earlier in this section. The main difference is that now the right hand side of \eqref{eq:rmrmr} becomes not a sum over an even and odd sector, but a sum over a contribution from all angular momenta appearing in all representations of $G$. Working this out in more detail (following e.g. \cite{Kapec:2019ecr}) gives an extra factor of $$2\sum_{J=0,\frac{1}{2},1,\ldots} (2J+1)^2, $$ multiplying the value of the ramp for a theory without any symmetry. The factor of $2$ comes from the ensemble being GOE/GSE while the $(2J+1)^2$ factor comes from the $\SU(2)$ structure.  This sum over spins gives a divergent factor (which can be regulated working in an ensemble where the angular momentum is restricted to a window $\delta J$ around an average value $J$) that is now at least qualitatively the same that we identified in the gravity calculation in Section \ref{sec:U1} and \ref{sec:SSSF}. We leave a more precise match of this normalization (involving a more careful analysis of quantum effects around the double cone in the presence of rotation) for future work.

\section{Discussion}\label{sec:discussion}

In this article we explained how to use the gravitational path integral to estimate the difference between the number of bosonic and fermionic black hole microstates. In particular, we focused on a contribution coming from the complex rotating black hole saddle point. This is a universal contribution that does not depend on the specific matter content of the theories, similar to the black hole that contributes to the ordinary partition function. We discussed its contribution for black holes in various dimensions, as well as the case of charged black holes in four dimensional flat space. We also describe wormhole contributions to the quantities we are computing similar to the double cone of \cite{Saad:2018bqo}. In this section we will make some further comments and point out some interesting directions. 

\paragraph{Purely bosonic theories} We stated at the beginning of Section \ref{sec:bh} that we are considering theories that contain fermions. However, one might be puzzled about where this assumption came into our discussion. Indeed, we never used the explicit details about the matter sector and only used the black hole geometries. On the other hand, our calculation must fail in a purely bosonic theory of quantum gravity (if it exists), since we are finding that the entropy with $(-1)^{\sf F}$ insertion is different from the entropy, meaning that there exists at least some fermionic states. 

In fact, the resolution to this puzzle is simple and instructive. In a theory with only bosons, turning on an angular potential $\Omega = 2\pi \i /\beta$ should be viewed as trivial, since all the states are invariant under a rotation by angle $2\pi$. However, this invariance is not explicit in terms of individual bulk saddle points. Instead, it is enforced by summing over saddles with shifts of $\Omega$ by $2\pi \i n /\beta, \, n\in \mathbb{Z}$. As a consequence, in both the calculation of $Z$ and $Z_{\textrm{spin}}$ we are summing over saddles with $\Omega = 2\pi \i \mathbb{Z}/\beta$, so we trivially have $Z = Z_{\textrm{spin}}$, consistent with the fact that the theory does not have fermions. In such theories, the rotating black holes analyzed in this paper are simply subleading contributions to the partition function.

So we see that our calculation relies on the assumption about the existence of fermions. In particular, it cannot be used to argue that a quantum gravity theory that contains black holes must have fermions. It would be nice to know whether this is true, and if so whether one can find some other argument for this statement. 

\paragraph{Other contributions to $\textrm{Tr}\, (-1)^{\sf F}$}
In this paper, we have only considered  universal contributions to the quantity $\textrm{Tr}\, (-1)^{\sf F}$, either in the canonical or microcanonical ensemble. Our result should really be thought of as giving a \emph{lower bound} to the size of this quantity, since in general there could be other more dominant contributions that depends on the details of the theory. (An upper bound, automatic from the boundary but not obvious from the bulk, is the partition function without the $(-1)^{\sf F}$.) One particular possibility is that a theory might contain codimension two defects which implements $(-1)^{\sf F}$ when one brings an operator around it. Such a defect can be placed at the tip of a Euclidean black hole such that it implements $(-1)^{\sf F}$ when going around the thermal circle, and it will give rise to a solution that computes $\textrm{Tr}\, (-1)^{\sf F}$.\footnote{We thank Miguel Montero for discussing this point.} This is a special case of the general story for black holes carrying discrete gauge charges \cite{Coleman:1991ku}. 

The detailed property of such a defect is a theory dependent question. However, there is an interesting construction of such a defect that only requires the knowledge of the low energy spectrum \cite{Arkani-Hamed:2007ryu}.\footnote{We thank Juan Maldacena for this suggestion.} One imagines that instead of having a Euclidean time circle which shrinks to zero at the horizon, it is stabilized to a finite radius by the Casimir energy coming from the light fields, in a similar way as described in \cite{Arkani-Hamed:2007ryu}. Given that the circle remains finite size, we are then free to choose the periodic boundary condition for the fermions, without the need to worry about the smoothness of the spin structure. Of course, this choice of boundary condition itself affects the Casimir energy one uses to find the solution. 

The intriguing aspect of such a solution is that it connects the properties of low energy spectrum to some properties of very high energy spectrum. We hope to return to the details of these solutions in the future.

\paragraph{Expectation from field theories} We studied a universal contribution from gravity to the difference between bosonic and fermionic states, using AdS/CFT. This raises the question of what expectations do we have from field theory for this quantity. 

The first obvious one is that the difference between bosonic and fermionic states cannot be large than the sum of them. This is obvious from field theory but becomes a non-trivial constraint from gravity. In particular, all the universal rotating black hole solutions we found satisfy this property. We found that $Z_\spin$, the partition function with a $(-1)^{\sf F}$ insertion, is exponentially subleading in the large $N$ limit compared to $Z$, the partition function without the insertion. This phenomenon was discussed in \cite{Cherman:2018mya} (see also references therein) for large $N$ adjoint QCD, where the authors find a large cancellation between bosonic and fermionic states. Our results were compared with theirs in the last paragraph of Section \ref{sec:ads4}.

A more interesting constraint comes from the thermal effective field theory put forth in \cite{Benjamin:2023qsc}. This predicts a specific dependence of the free energy with temperature and angular velocity. Their result was derived without the insertion of $(-1)^{\sf F}$. In the presence of such insertion one would expect their thermal effective theory still applies although the specific value of the effective theory parameters can change. \cite{Benjamin:2023qsc} also uses this free energy to extract the microcanonical density of states. Using the rotating black hole for the we find $F_\spin$ is controlled by thermal AdS, but in some cases like AdS${}_4$ the microcanonical density of states $S_\spin$ is controlled by the black hole. Furthermore, $S_{\spin}$ grows only as $\sqrt{E}$ and is subextensive in the volume ($S_{\textrm{spin}} \propto V^{\frac{3}{4}}$). It would be interesting to explore to what extent this is a violation of the thermal effective field theory.

\vskip.5cm
 \noindent {\it {Acknowledgements}}  
We thank Tom Hartman, Shota Komatsu, Henry Lin, Juan Maldacena, Miguel Montero, Sridip Pal, Douglas Stanford, Edward Witten for discussion. We specially thank Juan Maldacena for initial collaboration. YC is supported by a Procter Fellowship from Princeton University. GJTs work was supported by the Institute for Advanced Study and the NSF under Grant No. PHY-2207584, and by the Sivian Fund, and currently by the University of Washington and the DOE award DE-SC0024363.

 \appendix

\section{The Kontsevich-Segal-Witten criterion for complex Kerr saddles}\label{app:KSW}

In this appendix, we apply the Kontsevich-Segal-Witten (KSW) criterion for complex metrics \cite{Kontsevich:2021dmb,Witten:2021nzp} to the complex Kerr black hole solutions that are relevant for the discussion in this paper. The KSW criterion selects reasonable gravitational saddle points to be included in the gravitational path integral, which is derived by demanding that the fluctuations of various quantum fields on the background are suppressed. To what extent it is a strict rule that one should follow is still an open question.\footnote{For example, in the derivation it was assumed that the fluctuations of the quantum fields are integrated along real coutours. If one allows for deformations of the integration contours for the matter fields, then the criterion will be weakened. See also \cite{Maldacena:2019cbz,Bah:2022uyz} for some recently discussed geometries that violate the criterion but nonetheless lead to physically sensible results.} 

Concretely, the criterion states that for a complex metric $g$ on a $D$ dimensional manifold, if one picks a real basis such that the metric is diagonal
\begin{equation}
	g_{ij} = \lambda_i \delta_{ij}, \quad i,j = 1,...,D,
\end{equation}
then the KSW criterion demands that at every point of the manifold we have
\begin{equation}\label{eq:sumphase}
	\sum_{i=1}^D |\textrm{Arg}(\lambda_i)| < \pi. 
\end{equation}
In this appendix we discuss how the analysis can be done for either the Kerr metric in 4d flat space or Anti-de Sitter space. The discussion can be generalized other dimensions straightforwardly. 

The 4d Kerr black hole metric in flat space is 
\begin{equation}\label{eq:kerrallow}
\begin{aligned}
	\d s^2 = \frac{\rho^2 \Delta}{\Xi} \d t_E^2 + \frac{\rho^2}{\Delta} \d r^2 + \rho^2 \d \theta^2 + \frac{\Xi}{\rho^2} \sin^2 \theta\left( \d \tilde{\varphi} + \i \left(\frac{2Ear}{\Xi} - \Omega \right) \d t_E \right)^2 
\end{aligned}
\end{equation}
where the definition of various functions involved are given in (\ref{eq:Kerrdef}). Note that we've introduced coordinate $\tilde{\varphi}$ such that the coordinates are identified as $(t_E, \tilde{\varphi}) \sim (t_E + \beta, \tilde{\varphi}) \sim (t_E, \tilde{\varphi} + 2\pi)$. The metric is characterized by two parameters $(a,r_+)$. The case of $r_+,a \in \mathbb{R}$ was already analyzed in \cite{Witten:2021nzp}, with the conclusion being that the metric is disallowed under the criterion. The problem comes from that the imaginary piece in (\ref{eq:kerrallow}), which leads to a negative $g_{t_Et_E}$ component of the metric at large radius. The physical interpretation is simple - in flat space the fluctuations corresponding to adding particles far away from the black hole carrying large angular momenta are not suppressed. The situation becomes more complicated when we allow both $r_+$ and $a$ to be complex. In this case, it is not only the far away region we need to worry about, the near horizon region could also become dangerous. Even though the KSW criterion demands that (\ref{eq:sumphase}) should be satisfied everywhere on the manifold, the full analysis appears complicated and here we will only focus on the near horizon region as well as the asymptotic region, where the analysis can be simplified. We will see that simply from these two limits we can already get interesting constraints on $r_+,a$. 

We first focus on the near horizon part of the geometry. The analysis here is similar to the one for complexified non-rotating black holes in \cite{Chen:2022hbi}. By introducing new coordinates
\begin{equation}
	\d\varrho = \frac{\d r}{\sqrt{\Delta}}, \quad \d u = \frac{2\pi \d t_{E}}{\beta}, \quad u \sim u + 2\pi,
\end{equation}
the near horizon metric can be put into the following form
\begin{equation}
	\d s^2 = (r_+^2 + a^2 \cos^2 \theta) \left( \d \varrho^2 + \varrho^2 \d u^2 \right) + (r_+^2 + a^2 \cos^2 \theta) \d \theta^2 + \frac{r_+^2 + a^2}{r_+^2 + a^2 \cos^2 \theta} \sin^2 \theta \, \d \tilde{\varphi}^2 . 
\end{equation}
Note that the off-diagonal term $\d u \d \tilde{\varphi}$ will be of order $\varrho^2$ so can be ignored in the near horizon limit. We can absorb the factor $(r_+^2 + a^2 \cos^2 \theta)$ into $\d \varrho^2$ and therefore the radial plus time part of the metric can take a form that is completely real. 
As explained in \cite{Chen:2022hbi}, this is always achievable due to the smoothness of the geometry at the horizon. Therefore, the only nontrivial phases that will enter (\ref{eq:sumphase}) comes from $\d \theta^2$ and $\d \tilde{\varphi}^2$ parts of the metric. Concretely, (\ref{eq:sumphase}) requires
\begin{equation}
	\left| \textrm{Arg} \left( r_+^2 + a^2 \cos^2 \theta\right)\right| + \left| \textrm{Arg} \left( \frac{r_+^2 + a^2}{r_+^2 + a^2 \cos^2 \theta }\right)\right| < \pi
\end{equation}
for any $\theta\in [0,\pi)$. We could check that this is satisfied by the solutions we discussed in Section \ref{sec:4dflat}, since there we have $a = \i \zeta r_+$, with $r_+ \in \mathbb{R}$ and $\zeta$ taken to one from below.

We can also verify that our metric is allowable in the asymptotic region. The leading behavior of the metric at $r\rightarrow \infty$ takes the following form
\begin{equation}\label{eq:metinfinity}
	\d s^2 = - \frac{\beta^2}{4\pi^2} \Omega^2  r^2 \sin^2 \theta  \d u^2 - 2 \i r^2 \frac{\beta}{2\pi} \Omega \sin^2 \theta \d u \d \tilde{\varphi}  + \d r^2 + r^2 \d \theta^2 + r^2 \sin^2 \theta \d\tilde{\varphi}^2  .
\end{equation}
For the solution we considered in Section \ref{sec:4dflat}, we have $\beta \Omega = 2\pi \i$, which leads to a real metric in (\ref{eq:metinfinity}) with positive coefficients in front of $\d u^2$ and $\d \tilde{\varphi}^2$. From the general discussion in \cite{Witten:2021nzp} we know that such a metric is allowable. 

For the AdS$_4$ case discussed in Section \ref{sec:ads4}, the analysis is similar to the flat space case. The near horizon region imposes a family of constraints that says for any $\theta$ we should have
\begin{equation}\label{KSWads4}
	\left| \textrm{Arg} \left( \frac{r_+^2 + a^2 \cos^2 \theta }{\ell^2-a^2 \cos^2 \theta}\right)\right| + \left| \textrm{Arg} \left( \frac{(r_+^2 + a^2) (\ell^2 - a^2 \cos^2 \theta)}{(r_+^2 + a^2 \cos^2 \theta )(\ell^2 - a^2)^2 }\right)\right| < \pi .
\end{equation}
We've checked that the complex solution we considered in Section \ref{sec:ads4} satisfy (\ref{KSWads4}), while the solutions we dropped do not. Asymptotically, the metric (\ref{eq:ads4}) behaves as
\begin{equation}
	\d s^2 \approx  \frac{\ell^2}{r^2} \d r^2 +\frac{r^2}{\ell^2} \frac{\beta^2}{4\pi^2} \d u^2 + \i \frac{\beta}{2\pi}\frac{r^2}{\ell^2 } \frac{2a \sin^2 \theta}{1-a^2/\ell^2} \d u \d \phi +
 \frac{r^2 \sin^2\theta}{1-a^2/\ell^2} \d \phi^2 + \frac{r^2}{1 - \frac{a^2}{\ell^2} \cos^2 \theta} \d \theta^2 
\end{equation}
Since the metric contains off-diagonal terms, one has to first find a real basis in which the metric is diagonal and then apply the criterion (\ref{eq:sumphase}) \cite{Witten:2021nzp}. We performed this exercise and found that the solution we considered in (\ref{sec:ads4}) is also allowable in the asymptotic region.

\section{Smoothness of the Kerr black hole}\label{app:KerrR2}
In this section we explain how the Kerr black hole found in Section \ref{sec:4dflat} is smooth even though it requires taking a limit $r_+\to\infty$.  The curvature squared of the Kerr metric is given by 
\beq
R^{\mu\nu\rho\sigma}R_{\mu\nu\rho\sigma} = \frac{48 E^2 (r^2 - a^2 \cos^2\theta)[(r^2+a^2 \cos^2\theta)^2 - 16 r^2 a^2 \cos^2 \theta)]}{(r^2 + a^2 \cos^2 \theta)^6}.
\eeq
Since after setting $\beta \Omega $ to be pure imaginary the metric is real in Euclidean signature, we use a criterion for smoothness that this quantity is finite $|R^{\mu\nu\rho\sigma}R_{\mu\nu\rho\sigma}|< \infty$. The calculation is straightfoward so we will simply quote the result. As we take the $\varepsilon \to 0 $ limit of the solution in Section \ref{sec:4dflat} we find that for generic spacetime points the curvature squared vanishes $|R^{\mu\nu\rho\sigma}R_{\mu\nu\rho\sigma}|\sim \mathcal{O}(\varepsilon^6)$. The only exception is at the north and south pole $\theta = 0, \pi$  at the horizon $r=r_+$, where we find $|R^{\mu\nu\rho\sigma}R_{\mu\nu\rho\sigma}|= 24/E^4 +  \mathcal{O}(\varepsilon^1)$, which is still finite. 

\section{Review of spectral form factor in JT gravity with a $\U(1)$ gauge field}\label{app:JT}

Here we review the calculation of the spectral form factor in a random matrix model with $\U(1)$ global symmetry, dual to JT gravity coupled to a $\U(1)$ Maxwell theory. The partition functions for general Riemann surfaces for this theory was discussed in \cite{Kapec:2019ecr}. Here we review their story for the two boundary wormhole, and in particular, the computation where we focus on a microcanonical window around energy $E$, with width $\Delta$. We will work in the unit that the length scale $\phi_r$ governing the Schwarzian fluctuations in JT is one.

We start with the expression for the double trumpet in this theory \cite{Kapec:2019ecr}
\begin{equation}\label{double}
	Z(\beta_L,\mu_L;\beta_R,\mu_R) = 2\pi \int_0^{2\pi} \d\phi \int_0^\infty \d b\,b\, Z_{tr}(\beta_L,\mu_L;b,\phi) Z_{tr}(\beta_R,\mu_R;b,-\phi),
\end{equation}
where $Z_{tr} (\beta, \mu;b,\phi) = Z_{tr}^{gauge} Z_{tr}^{grav}$, with $Z_{tr}^{grav}$ being the ordinary JT trumpet partition function
\begin{equation}
	Z_{tr}^{grav}(\beta) = \frac{1}{\sqrt{2\pi \beta }} \exp\left[- \frac{b^2}{2\beta}\right]
\end{equation}
and 
\begin{equation}\label{Ztrgauge}
\begin{aligned}
	Z_{tr}^{gauge}(\beta,\mu,\phi) & = \frac{1}{\sqrt{4\pi \beta}} \sum_{n \in \mathbb{Z}} \exp\left[ - \frac{1}{2\beta} (2\pi n - \i \mu \beta - \phi)^2 \right]\\
	& = \frac{1}{\sqrt{8\pi^2}} \sum_{n\in \mathbb{Z}}\int \d q\, \exp\left[q (\beta\mu + 2\pi \i n - \i \phi)\right]\exp\left[ - \beta \frac{q^2}{2}\right].
\end{aligned}
\end{equation} 
The variable $\phi$ in (\ref{double}) and (\ref{Ztrgauge}) comes from the holonomy of the gauge field at the throat of the wormhole. Note that in the second line of (\ref{Ztrgauge}), we could further sum over $n$ which sets $q$ to be integers, but we will defer this step such that the procedure is more analogues to the general story described in Section \ref{sec:U1}. Taking (\ref{Ztrgauge}) into (\ref{double}), we get
\begin{equation}\label{Zdouble}
\begin{aligned}
	Z(\beta_L,\mu_L;\beta_R,\mu_R) = & \frac{1}{8\pi^2} \sum_{n_L,n_R} \int \d q_L \int \d q_R \int_0^{2\pi} \d\phi  \int_0^\infty \d b\, b \frac{1}{\sqrt{\beta_L\beta_R}} \exp\left[ - \frac{b^2}{2\beta_L} -  \frac{b^2}{2\beta_R} \right]\\
	& \exp\left[ \beta_L\mu_L q_L + \beta_R \mu_R q_R \right]\exp\left[ - \beta_L \frac{q_L^2}{2}-\beta_R \frac{q_R^2}{2}\right]\\
	& \exp\left[q_L (2\pi \i n_L - \i \phi)+q_R ( 2\pi \i n_R + \i \phi)\right]
\end{aligned}
\end{equation}
Redefine $n= n_R ,m = n_L + n_R$, we can rewrite the exponent on the last line of (\ref{Zdouble}) as
\begin{equation}
	q_L \left(2\pi \i (m - n) - \i\phi\right) + q_R\left(2\pi \i n + \i\phi\right) = 2\pi \i m q_L + (2\pi \i (n+\phi)) (q_R-q_L).
\end{equation}
We can combine the sum over $n$ and the integral over $\phi$ into a single integral of $n+\phi$ over the real axis, which imposes that $q_L= q_R$. Then we have
\begin{equation}
\begin{aligned}
	Z(\beta_L,\mu_L;\beta_R,\mu_R) = & \frac{1}{8\pi^2} \sum_{m} \int \d q \int_0^\infty \d b\, b \frac{1}{\sqrt{\beta_L\beta_R}} \exp\left[ - \frac{b^2}{2\beta_L} -  \frac{b^2}{2\beta_R} \right]\\
	& \exp\left[ q (\beta_L\mu_L +\beta_R \mu_R  + 2\pi \i m )  \right]\exp\left[ - \beta_L \frac{q^2}{2}-\beta_R \frac{q^2}{2}\right]
\end{aligned}
\end{equation}
Again, we defer the sum over $m$ which sets the charge to be integers.

We are interested in the quantity $\langle |Y_{E,\mu}(T)|^2\rangle_c$ defined in (\ref{Ycharge}). It can be computed by applying a transformation of (\ref{double})
\begin{equation}\label{Y2append}
\begin{aligned}
	\langle |Y_{E,\mu}(T)|^2\rangle_c  & \propto  \int \d\beta_L \d\beta_R  \, e^{\beta_L E + \beta_R E + \frac{1}{2} (\beta_L^2 + \beta_R^2)\Delta^2}Z(\beta_L-\i T,\mu;\beta_R+\i T,\mu) \\
	& \propto  \int \d\beta_L \d\beta_R  \, e^{\beta_L E + \beta_R E + \frac{1}{2} (\beta_L^2 + \beta_R^2)\Delta^2} \times \\
	& \quad \sum_{m} \int \d q \int_0^\infty \d b\, b \frac{1}{\sqrt{(\beta_L-\i T)(\beta_R+\i T)}} \exp\left[ - \frac{b^2}{2(\beta_L-\i T)} -  \frac{b^2}{2(\beta_R+\i T)} \right]\\
	& \quad \exp\left[ q ((\beta_L+\beta_R)\mu   + 2\pi \i m )  \right]\exp\left[ - (\beta_L-\i T) \frac{q^2}{2}-(\beta_R + \i T) \frac{q^2}{2}\right]
\end{aligned}
\end{equation}
To understand this seemingly complicated expression, we can look for the saddle points for the integrals for $\beta_L,\beta_R$ and $b$. Similar to the ordinary double cone \cite{Saad:2018bqo}, the saddle point value of $\beta_L$ and $\beta_R$ are located at zero, while the saddle point for $b$ is located at
\begin{equation}
	b_* = \sqrt{2\left(E + \mu q - \frac{q^2}{2}\right)} T.
\end{equation}
We can then expand (\ref{Y2append}) around the saddle points and evaluate the one loop determinant. The final result of the calculation is very simple
\begin{equation}\label{Yfinal}
\begin{aligned}
	\langle |Y_{E,\mu}(T)|^2\rangle_c & \propto  \sum_m \int \d q \, e^{2\pi \i m q} \, T  \\
	&  \propto \sum_q T.
\end{aligned}
\end{equation}
We kept the sum over $m$ in the first line because in the discussion of \ref{sec:U1} we did not discuss the effect of summing over shifts (we had $m=0$ there), so we simply have a continuous integral over the $\U(1)$ charge. Of course, the effect of summing over $m$ is to enforce charge quantization. The final result (\ref{Yfinal}) is easy to interpret. We simply have an independent random matrix in each of the charge sectors, which contributes one copy of the linear $T$ growth. The final result is the sum over all the charge sectors. In particular, the final result is independent of the chemical potential $\mu$.

\bibliographystyle{JHEP}
\bibliography{bib}

\providecommand{\href}[2]{#2}\begingroup\raggedright\begin{thebibliography}{10}

\bibitem{Gibbons:1976ue}
G.~W. Gibbons and S.~W. Hawking, \emph{{Action Integrals and Partition
  Functions in Quantum Gravity}},
  \href{https://doi.org/10.1103/PhysRevD.15.2752}{\emph{Phys. Rev. D}
  {\bfseries 15} (1977) 2752}.

\bibitem{Cherman:2018mya}
A.~Cherman, M.~Shifman and M.~\"Unsal, \emph{{Bose-Fermi cancellations without
  supersymmetry}},
  \href{https://doi.org/10.1103/PhysRevD.99.105001}{\emph{Phys. Rev. D}
  {\bfseries 99} (2019) 105001}
  [\href{https://arxiv.org/abs/1812.04642}{{\ttfamily 1812.04642}}].

\bibitem{Cabo-Bizet:2018ehj}
A.~Cabo-Bizet, D.~Cassani, D.~Martelli and S.~Murthy, \emph{{Microscopic origin
  of the Bekenstein-Hawking entropy of supersymmetric AdS$_{5}$ black holes}},
  \href{https://doi.org/10.1007/JHEP10(2019)062}{\emph{JHEP} {\bfseries 10}
  (2019) 062} [\href{https://arxiv.org/abs/1810.11442}{{\ttfamily
  1810.11442}}].

\bibitem{Bobev:2020pjk}
N.~Bobev, A.~M. Charles and V.~S. Min, \emph{{Euclidean black saddles and
  AdS$_{4}$ black holes}},
  \href{https://doi.org/10.1007/JHEP10(2020)073}{\emph{JHEP} {\bfseries 10}
  (2020) 073} [\href{https://arxiv.org/abs/2006.01148}{{\ttfamily
  2006.01148}}].

\bibitem{Iliesiu:2021are}
L.~V. Iliesiu, M.~Kologlu and G.~J. Turiaci, \emph{{Supersymmetric indices
  factorize}}, \href{https://doi.org/10.1007/JHEP05(2023)032}{\emph{JHEP}
  {\bfseries 05} (2023) 032}
  [\href{https://arxiv.org/abs/2107.09062}{{\ttfamily 2107.09062}}].

\bibitem{Witten:1998zw}
E.~Witten, \emph{{Anti-de Sitter space, thermal phase transition, and
  confinement in gauge theories}},
  \href{https://doi.org/10.4310/ATMP.1998.v2.n3.a3}{\emph{Adv. Theor. Math.
  Phys.} {\bfseries 2} (1998) 505}
  [\href{https://arxiv.org/abs/hep-th/9803131}{{\ttfamily hep-th/9803131}}].

\bibitem{Cotler:2016fpe}
J.~S. Cotler, G.~Gur-Ari, M.~Hanada, J.~Polchinski, P.~Saad, S.~H. Shenker
  et~al., \emph{{Black Holes and Random Matrices}},
  \href{https://doi.org/10.1007/JHEP05(2017)118}{\emph{JHEP} {\bfseries 05}
  (2017) 118} [\href{https://arxiv.org/abs/1611.04650}{{\ttfamily
  1611.04650}}].

\bibitem{Saad:2018bqo}
P.~Saad, S.~H. Shenker and D.~Stanford, \emph{{A semiclassical ramp in SYK and
  in gravity}},  \href{https://arxiv.org/abs/1806.06840}{{\ttfamily
  1806.06840}}.

\bibitem{Pal:2020wwd}
S.~Pal and Z.~Sun, \emph{{High Energy Modular Bootstrap, Global Symmetries and
  Defects}}, \href{https://doi.org/10.1007/JHEP08(2020)064}{\emph{JHEP}
  {\bfseries 08} (2020) 064}
  [\href{https://arxiv.org/abs/2004.12557}{{\ttfamily 2004.12557}}].

\bibitem{Harlow:2021trr}
D.~Harlow and H.~Ooguri, \emph{{A universal formula for the density of states
  in theories with finite-group symmetry}},
  \href{https://doi.org/10.1088/1361-6382/ac5db2}{\emph{Class. Quant. Grav.}
  {\bfseries 39} (2022) 134003}
  [\href{https://arxiv.org/abs/2109.03838}{{\ttfamily 2109.03838}}].

\bibitem{Mukhametzhanov:2020swe}
B.~Mukhametzhanov and S.~Pal, \emph{{Beurling-Selberg Extremization and Modular
  Bootstrap at High Energies}},
  \href{https://doi.org/10.21468/SciPostPhys.8.6.088}{\emph{SciPost Phys.}
  {\bfseries 8} (2020) 088} [\href{https://arxiv.org/abs/2003.14316}{{\ttfamily
  2003.14316}}].

\bibitem{Pal:2022vqc}
S.~Pal, J.~Qiao and S.~Rychkov, \emph{{Twist Accumulation in Conformal Field
  Theory: A Rigorous Approach to the Lightcone Bootstrap}},
  \href{https://doi.org/10.1007/s00220-023-04767-w}{\emph{Commun. Math. Phys.}
  {\bfseries 402} (2023) 2169}
  [\href{https://arxiv.org/abs/2212.04893}{{\ttfamily 2212.04893}}].

\bibitem{Pal:2023cgk}
S.~Pal and J.~Qiao, \emph{{Lightcone Modular Bootstrap and Tauberian Theory: A
  Cardy-like Formula for Near-extremal Black Holes}},
  \href{https://arxiv.org/abs/2307.02587}{{\ttfamily 2307.02587}}.

\bibitem{Witten:1982df}
E.~Witten, \emph{{Constraints on Supersymmetry Breaking}},
  \href{https://doi.org/10.1016/0550-3213(82)90071-2}{\emph{Nucl. Phys. B}
  {\bfseries 202} (1982) 253}.

\bibitem{H:2023qko}
A.~A. H., P.~V. Athira, C.~Chowdhury and A.~Sen, \emph{{Logarithmic Correction
  to BPS Black Hole Entropy from Supersymmetric Index at Finite Temperature}},
  \href{https://arxiv.org/abs/2306.07322}{{\ttfamily 2306.07322}}.

\bibitem{Witten:2021nzp}
E.~Witten, \emph{{A Note On Complex Spacetime Metrics}},
  \href{https://arxiv.org/abs/2111.06514}{{\ttfamily 2111.06514}}.

\bibitem{Ghosh:2019rcj}
A.~Ghosh, H.~Maxfield and G.~J. Turiaci, \emph{{A universal Schwarzian sector
  in two-dimensional conformal field theories}},
  \href{https://arxiv.org/abs/1912.07654}{{\ttfamily 1912.07654}}.

\bibitem{Iliesiu:2020qvm}
L.~V. Iliesiu and G.~J. Turiaci, \emph{{The statistical mechanics of
  near-extremal black holes}},
  \href{https://doi.org/10.1007/JHEP05(2021)145}{\emph{JHEP} {\bfseries 05}
  (2021) 145} [\href{https://arxiv.org/abs/2003.02860}{{\ttfamily
  2003.02860}}].

\bibitem{Heydeman:2020hhw}
M.~Heydeman, L.~V. Iliesiu, G.~J. Turiaci and W.~Zhao, \emph{{The statistical
  mechanics of near-BPS black holes}},
  \href{https://doi.org/10.1088/1751-8121/ac3be9}{\emph{J. Phys. A} {\bfseries
  55} (2022) 014004} [\href{https://arxiv.org/abs/2011.01953}{{\ttfamily
  2011.01953}}].

\bibitem{Turiaci:2023wrh}
G.~J. Turiaci, \emph{{New insights on near-extremal black holes}},
  \href{https://arxiv.org/abs/2307.10423}{{\ttfamily 2307.10423}}.

\bibitem{Chamblin:1999tk}
A.~Chamblin, R.~Emparan, C.~V. Johnson and R.~C. Myers, \emph{{Charged AdS
  black holes and catastrophic holography}},
  \href{https://doi.org/10.1103/PhysRevD.60.064018}{\emph{Phys. Rev.}
  {\bfseries D60} (1999) 064018}
  [\href{https://arxiv.org/abs/hep-th/9902170}{{\ttfamily hep-th/9902170}}].

\bibitem{Kontsevich:2021dmb}
M.~Kontsevich and G.~Segal, \emph{{Wick Rotation and the Positivity of Energy
  in Quantum Field Theory}},
  \href{https://doi.org/10.1093/qmath/haab027}{\emph{Quart. J. Math. Oxford
  Ser.} {\bfseries 72} (2021) 673}
  [\href{https://arxiv.org/abs/2105.10161}{{\ttfamily 2105.10161}}].

\bibitem{Maldacena:1998bw}
J.~M. Maldacena and A.~Strominger, \emph{{AdS(3) black holes and a stringy
  exclusion principle}},
  \href{https://doi.org/10.1088/1126-6708/1998/12/005}{\emph{JHEP} {\bfseries
  12} (1998) 005} [\href{https://arxiv.org/abs/hep-th/9804085}{{\ttfamily
  hep-th/9804085}}].

\bibitem{Maloney:2007ud}
A.~Maloney and E.~Witten, \emph{{Quantum Gravity Partition Functions in Three
  Dimensions}}, \href{https://doi.org/10.1007/JHEP02(2010)029}{\emph{JHEP}
  {\bfseries 02} (2010) 029} [\href{https://arxiv.org/abs/0712.0155}{{\ttfamily
  0712.0155}}].

\bibitem{MYITP:2023sep}
D.~Harlow,
  \emph{{\href{https://www2.yukawa.kyoto-u.ac.jp/~qimg2023/week2.php}{Gauging
  spacetime inversions}}}, {\emph{{Talk at ExU-YITP Workshop on Holography,
  Gravity and Quantum Information}} (2023) }.

\bibitem{Hartman:2014oaa}
T.~Hartman, C.~A. Keller and B.~Stoica, \emph{{Universal Spectrum of 2d
  Conformal Field Theory in the Large c Limit}},
  \href{https://doi.org/10.1007/JHEP09(2014)118}{\emph{JHEP} {\bfseries 09}
  (2014) 118} [\href{https://arxiv.org/abs/1405.5137}{{\ttfamily 1405.5137}}].

\bibitem{Penington:2019kki}
G.~Penington, S.~H. Shenker, D.~Stanford and Z.~Yang, \emph{{Replica wormholes
  and the black hole interior}},
  \href{https://arxiv.org/abs/1911.11977}{{\ttfamily 1911.11977}}.

\bibitem{Stanford:2020wkf}
D.~Stanford, \emph{{More quantum noise from wormholes}},
  \href{https://arxiv.org/abs/2008.08570}{{\ttfamily 2008.08570}}.

\bibitem{Choi:2022asl}
S.~Choi, S.~Kim and J.~Song, \emph{{Supersymmetric Spectral Form Factor and
  Euclidean Black Holes}},  \href{https://arxiv.org/abs/2206.15357}{{\ttfamily
  2206.15357}}.

\bibitem{Kapec:2019ecr}
D.~Kapec, R.~Mahajan and D.~Stanford, \emph{{Matrix ensembles with global
  symmetries and \textquoteright{}t Hooft anomalies from 2d gauge theory}},
  \href{https://doi.org/10.1007/JHEP04(2020)186}{\emph{JHEP} {\bfseries 04}
  (2020) 186} [\href{https://arxiv.org/abs/1912.12285}{{\ttfamily
  1912.12285}}].

\bibitem{Cotler:2022rud}
J.~Cotler and K.~Jensen, \emph{{A precision test of averaging in AdS/CFT}},
  \href{https://doi.org/10.1007/JHEP11(2022)070}{\emph{JHEP} {\bfseries 11}
  (2022) 070} [\href{https://arxiv.org/abs/2205.12968}{{\ttfamily
  2205.12968}}].

\bibitem{CIM}
Y.~Chen, V.~Ivo and J.~Maldacena, \emph{{Comments on the double cone
  wormhole}}, {\emph{to appear} }.

\bibitem{Stanford:2019vob}
D.~Stanford and E.~Witten, \emph{{JT gravity and the ensembles of random matrix
  theory}}, \href{https://doi.org/10.4310/ATMP.2020.v24.n6.a4}{\emph{Adv.
  Theor. Math. Phys.} {\bfseries 24} (2020) 1475}
  [\href{https://arxiv.org/abs/1907.03363}{{\ttfamily 1907.03363}}].

\bibitem{Yan:2023rjh}
C.~Yan, \emph{{More on Torus Wormholes in 3d Gravity}},
  \href{https://arxiv.org/abs/2305.10494}{{\ttfamily 2305.10494}}.

\bibitem{Witten:2023snr}
E.~Witten, \emph{{Anomalies and Nonsupersymmetric D-Branes}},
  \href{https://arxiv.org/abs/2305.01012}{{\ttfamily 2305.01012}}.

\bibitem{Coleman:1991ku}
S.~R. Coleman, J.~Preskill and F.~Wilczek, \emph{{Quantum hair on black
  holes}}, \href{https://doi.org/10.1016/0550-3213(92)90008-Y}{\emph{Nucl.
  Phys. B} {\bfseries 378} (1992) 175}
  [\href{https://arxiv.org/abs/hep-th/9201059}{{\ttfamily hep-th/9201059}}].

\bibitem{Arkani-Hamed:2007ryu}
N.~Arkani-Hamed, S.~Dubovsky, A.~Nicolis and G.~Villadoro, \emph{{Quantum
  Horizons of the Standard Model Landscape}},
  \href{https://doi.org/10.1088/1126-6708/2007/06/078}{\emph{JHEP} {\bfseries
  06} (2007) 078} [\href{https://arxiv.org/abs/hep-th/0703067}{{\ttfamily
  hep-th/0703067}}].

\bibitem{Benjamin:2023qsc}
N.~Benjamin, J.~Lee, H.~Ooguri and D.~Simmons-Duffin, \emph{{Universal
  Asymptotics for High Energy CFT Data}},
  \href{https://arxiv.org/abs/2306.08031}{{\ttfamily 2306.08031}}.

\bibitem{Maldacena:2019cbz}
J.~Maldacena, G.~J. Turiaci and Z.~Yang, \emph{{Two dimensional Nearly de
  Sitter gravity}}, \href{https://doi.org/10.1007/JHEP01(2021)139}{\emph{JHEP}
  {\bfseries 01} (2021) 139}
  [\href{https://arxiv.org/abs/1904.01911}{{\ttfamily 1904.01911}}].

\bibitem{Bah:2022uyz}
I.~Bah, Y.~Chen and J.~Maldacena, \emph{{Estimating global charge violating
  amplitudes from wormholes}},
  \href{https://doi.org/10.1007/JHEP04(2023)061}{\emph{JHEP} {\bfseries 04}
  (2023) 061} [\href{https://arxiv.org/abs/2212.08668}{{\ttfamily
  2212.08668}}].

\bibitem{Chen:2022hbi}
Y.~Chen, \emph{{Spectral form factor for free large N gauge theory and
  strings}}, \href{https://doi.org/10.1007/JHEP06(2022)137}{\emph{JHEP}
  {\bfseries 06} (2022) 137}
  [\href{https://arxiv.org/abs/2202.04741}{{\ttfamily 2202.04741}}].

\end{thebibliography}\endgroup
\end{document}